\documentclass[a4paper,openright,11pt]{article}
\usepackage{fancyhdr}
\usepackage{graphicx} 
\usepackage{amsfonts}
\usepackage{amsmath,amsthm,amssymb}
\usepackage{mathptmx}

\usepackage[utf8]{inputenc}
\usepackage{dsfont}
\usepackage[english]{babel}
\usepackage{cite}
\usepackage{subfigure} 
\usepackage[hyperindex=true,breaklinks=true,colorlinks=true,linkcolor=blue,citecolor=blue]{hyperref}

\numberwithin{equation}{section}

\title{Analysis of a one-dimensional Hamiltonian with a singular double well consisting of two nonlocal $\delta'$ interactions}

\author{
S. Fassari$^{1}$\footnote{sifassari@gmail.com, 
ORCID: \href{http://orcid.org/0000-0003-3475-7696}{0000-0003-3475-7696}}, 
M. Gadella$^2$\footnote{manuelgadella1@gmail.com, 
ORCID: \href{http://orcid.org/0000-0001-8860-990X}{0000-0001-8860-990X}}, 
L. M. Nieto$^2$\footnote{luismiguel.nieto.calzada@uva.es, ORCID: \href{http://orcid.org/0000-0002-2849-2647}{0000-0002-2849-2647}},
F. Rinaldi$^{1}$\footnote{f.rinaldi@unimarconi.it, 
ORCID: \href{http://orcid.org/0000-0002-0087-3042}{0000-0002-0087-3042}}
\medskip
\\ 
\noindent
\small
\noindent
$^1$\,Dipartimento di Science Ingegneristiche. Univ. degli Studi Guglielmo Marconi,\\
\small
Via Plinio 44, I-00193 Rome, Italy\\
\small
\noindent
$^2$\,Departamento de F\'{\i}sica Te\'orica, At\'omica y
\'Optica,
Universidad de Valladolid,  47011 Valladolid, Spain
}

\date{\today}

\begin{document}

\maketitle

\begin{abstract}
The objective of the present paper is the study of a one-dimensional Hamiltonian with the interaction term given by the sum of two nonlocal attractive $\delta'$-interactions of equal strength and symmetrically located with respect to the origin. We use the procedure known as {\it renormalisation of the coupling constant} in order to rigorously achieve a self-adjoint determination for this Hamiltonian. This model depends on two parameters, the interaction strength and the distance between the centre of each interaction and the origin. Once we have the self-adjoint determination, we obtain its discrete spectrum showing that it consists of two negative eigenvalues representing the energy levels. We analyse the dependence of these energy levels on the above-mentioned parameters. We investigate the possible resonances of the model. Furthermore, we analyse in detail the limit of our model as the distance between the supports of the two $\delta'$ interactions vanishes.
\end{abstract}

\section{Introduction}

This is a new contribution to the study of one-dimensional contact potentials, or potentials with support consisting of a single point or a discrete collection of points \cite{AGHH,AK,BR,DO,FGGN,ZZ,ZZ1,FI,FI1,Z}. There are two main reasons for the study of this type of objects. From a physicist's point of view, one-dimensional Hamiltonians with contact interactions are used to model a wide range of situations. For instance, those including extra thin structures, point defects in materials, heterostructures with abrupt effective mass change, in addition to other applications in the study of nanostructures. They also provide one-particle states in scalar $(1+1)$-dimensional QFT, Casimir effect, etc \cite{MMM,AGM,AM,MKB,MGMC,BM,BOR,UTM,APST,DE,GMMN,ZZ2}. In addition, many one-dimensional models with two or more contact potentials show scattering resonances and other scattering features, as poles of the analytically continued $S$-matrix (or reduced resolvent), thus being a useful source for the study of unstable quantum systems \cite{Z1,Z2,EK,EGU,EGU1,EGU2,AGLM}. 

From the mathematical point of view, contact potentials appear in the theory of self-adjoint extensions of symmetric operators. In this case, each self-adjoint extension with a contact potential supported at {\it one point} is characterised by some conditions that must be satisfied by the functions belonging to its domain on the support of the contact potential. Then, we may characterise the potential by one of these conditions. Nevertheless, there are particular situations in which the determination of such constraints for a given predetermined contact potential is not easy. Instead, we have to resort to other strategies in order to provide a self-adjoint determination to the given formal Hamiltonian. These strategies often require a renormalisation and the resulting self-adjoint Hamiltonian is defined via its resolvent or its Birman-Schwinger operator which, in some sense, gives a shortcut to the problem of finding eigenvalues and resonances arising as a result of the renormalised potential.  

One-dimensional nonrelativistic contact potentials having support at one single point have been classified in \cite{KU}. In this case, one gets four one-dimensional families of self-adjoint extensions of the symmetric operator $ H_0= -d^2/dx^2$ on a suitable domain in $L^2(\mathbb R)$, where each extension is readily characterised by two-sided boundary conditions on the wave functions of the domain of the extension on the support of the potential. Physical interpretations of the resulting potentials have been given \cite{KP,KP1}, even though the general consensus on them is still far from being achieved. Contact potentials perturbing the Salpeter Hamiltonian $\sqrt{-d^2/dx^2 + m^2}$, which, differently from the Laplacian, is characterised by being {\it nonlocal}, have been studied in the literature. Here, there is a unique family of point perturbations, given by $\alpha\delta(x)$ with $\alpha \in \mathbb R$, that makes the total Hamiltonian self-adjoint, so that $H=\sqrt{-d^2/dx^2 + m^2} + \alpha \, \delta(x)$. In this case, self-adjoint determinations for each value of the parameter $\alpha$ cannot be given by matching conditions, which implies that the renormalisation procedure is required \cite{AK1,HSW,HS,EGU}, a feature characterising also Hamiltonians with contact potentials supported at {\it one point} in two or three dimensions and some others \cite{FI2,BG,AFR,AFR1,E,EST,LL}. 

Both renormalisation and the construction of Birman-Schwinger formulae \cite{K,K1} may represent a mathematical challenge that makes the procedure interesting from the mathematical point of view. 

With regard to the unperturbed Hamiltonian $H_0= -d^2/dx^2$, a typical domain $\mathcal D(H_0)$ on which $H_0$ is symmetric is given by ($f(x)$ is a measurable function $f(x): \mathbb R \longmapsto \mathbb C$ with properties as below):
\begin{equation}\label{1}
\mathcal D(H_0) := \{ f(x) \in W^2_2 (\mathbb R)\,, \;\;\; f(x_0) = f'(x_0) =0\}\,,
\end{equation}
for some fixed $x_0 \in \mathbb R$ (often $x_0=0$). Here $W^2_2(\mathbb R)$ is the Sobolev space of absolutely continuous square integrable functions, $f(x)$, on $\mathbb R$ having an absolutely continuous first derivative and a square integrable second derivative, so that 
\begin{equation}\label{2}
\int_{-\infty}^\infty \left(|f(x)|^2 + |f''(x)|^2\right)\, dx <\infty\,.
\end{equation}

In the present paper we study a one-dimensional Hamiltonian decorated with two attractive nonlocal $\delta'$-interactions symmetrically located around the origin, so that we may start from the merely heuristic expression:
\begin{equation}\label{3}
H(\lambda,x_0) = - \frac{d^2}{dx^2} -\lambda \left[ \delta'(x+x_0) + \delta'(x-x_0)\right]\,, \qquad \lambda>0\,.
\end{equation}

The term {\it nonlocal}, which has been used in previous papers by the authors, see for instance \cite{FGGN}, requires an explanation. In fact, there exist two possible $\delta'$-interactions, one local and the other nonlocal. 

To begin with, let us recall that the $\delta'$, regarded as a distribution, is defined by
$$
(\delta',f)=-(\delta,f')=-f'(0),\,   f\in {\mathcal S}(\mathbb R),
$$
where ${\mathcal S}(\mathbb R)$ is the well-known Schwartz space of test functions, the properties of which have been presented in detail in \cite{RSI,Scharf}.

In the theory of distributions one often wishes to assess how singular a given distribution is. The concept playing a crucial role in this assessment is the so-called {\it singular order of a distribution}. Although its rigorous definition was given in \cite{Scharf} both in coordinate and momentum space, we wish to recall here the more operational criterion provided in \cite{Tadeu} in order to determine the singular order of a given distribution. A distribution (with respect to any space of one-dimensional test functions) $T$ has singular order equal to $s$ if $T$ is the $(s+2)$-th derivative in the sense of distributions of a given continuous (not necessarily differentiable) function $f(x)$, so that $T= D^{s+2}f$, where $D$ means derivative in the distributional sense. Here $s$ is any integer and a derivative of negative order denotes an indefinite integral.  Thus, the Heaviside function, regarded as a distribution, has singular order $s=-1$ (it is not continuous, even though it is the distributional derivative of a  continuous function). Therefore, its first derivative, the Dirac distribution $\delta$, has singular order $s=0$, while its second derivative $\delta'$ has singular order $s=1$.

As was mentioned earlier, two different point perturbations of the one-dimensional Laplacian, both stemming from the $\delta'$-distribution, have been studied in the literature. The former, called {\it local}  $\delta'$-interaction or $\delta'$-potential at $x_0$ , denoted by  $\delta'(x-x_0)$, is defined by its action on a pair of real valued test functions $f(x)$ and $g(x)$ (we use real valued functions in order to make the notation lighter, the extension to complex valued functions being quite obvious):
\begin{eqnarray}
(g, \delta'(x-x_0)\, f) \!\!&\!\!=\!\!&\!\!  \int_{\infty}^\infty g(x)\, \delta'(x-x_0)\, f(x)\, dx = -  \int_{\infty}^\infty \delta(x-x_0) \, \frac {d}{dx}\, [g(x)f(x)]\, dx \nonumber  \\
 \!\!&\!\!=\!\!&\!\!  -g(x_0)\, f'(x_0)- g'(x_0)\, f(x_0).
\label{4}
\end{eqnarray}

The latter, called {\it nonlocal} $\delta'$-interaction at $x_0$, acts as a dyad of the form $ |\delta'_{x_0}\rangle \langle \delta'_{x_0}|$, so that for any (real) test functions $f(x)$ and $g(x)$, we have:
 \begin{eqnarray}
(g, |\delta'_{x_0}\rangle\langle \delta'_{x_0}|\, f) \!\!&\!\!=\!\!&\!\!  (g, \delta'(x-x_0) )\ ( \delta'(x-x_0),f) = 
[-(g',\delta(x-x_0))]\, [-(\delta(x-x_0),f')] \nonumber  \\
 \!\!&\!\!=\!\!&\!\! g'(x_0)\, f'(x_0).
\label{5}
\end{eqnarray}

It should be remarked that such a distinction does not exist in the case of the Dirac delta. In fact, if 
$ \delta(x-x_0)$ is the Dirac delta at $x_0$, we have for any real test functions $f(x)$ and $g(x)$ 
\begin{eqnarray}
(g,\delta(x-x_0) \, f) \!\!&\!\!=\!\!&\!\!  \int_{\infty}^\infty g(x) \, \delta(x-x_0) \, f(x)\, dx = g(x_0)\, f(x_0) 
\nonumber  \\
 \!\!&\!\!=\!\!&\!\! (g,\delta(x-x_0) ) (\delta(x-x_0)  ,f) = (g, \, |\delta_{x_0}\rangle\langle \delta_{x_0} |\, f)\,.
\label{6}
\end{eqnarray}

As is well known, the spaces $\mathcal H_n$ are the spaces of measurable functions $f(x):\mathbb R \longmapsto \mathbb C$ such that
\begin{equation}\label{7}
\int_{-\infty}^\infty (1+p^2)^{n}\,|\hat f(p)|^2\,  dp= \|(1+p^2)^{\frac {n}2}\,\hat f\|_2^2 <\infty\,.
\end{equation}

The dual of $\mathcal H_n$ is $\mathcal H_{-n}$. Then, while $\delta\in \mathcal H_{-1}$ due to the renowned KLMN theorem (see \cite{RSII,DONSIMON}), the nonlocal $\delta' \in \mathcal H_{-2}$ since for any $f \in \mathcal H_{2}$:
\begin{equation}
(f, |\delta'><\delta'| f)=(f', \delta f'),
\end{equation}
so that the KLMN theorem is applicable to  $f' \in \mathcal H_{1}$.

Incidentally, the fact that $|\delta'><\delta'|=\frac {d}{dx}\,\delta \frac {d}{dx}$ explains why the term "momentum dependent interaction" was coined for the nonlocal $\delta'$-interaction in the literature (see \cite{GLP,AFR4}).

Regarded as perturbations of $H_0=-d^2/dx^2$, the point potentials given by the Dirac delta or either the local or the nonlocal $\delta'$, determine self-adjoint Hamiltonians. These self-adjoint determinations are given by their respective domains characterised by  two-sided boundary conditions at the point supporting the point potentials. 
As is well known (see page 157 in \cite{AK}), the  two-sided boundary conditions for any function $\psi(x)$ in the domain of $H_0 +
 a\, \delta(x)$ are \cite{GMMN}
\begin{equation}\label{8}
\left(\begin{array}{c} \psi(+0) \\ \psi'(+0) \end{array}  \right) =  \left(\begin{array}{cc} 1 & 0 \\ a & 1 \end{array}  \right)
 \left(\begin{array}{c}  \psi(-0)  \\ \psi'(-0)  \end{array}  \right),
\end{equation}
while the  two-sided boundary conditions  for the Hamiltonian $H_0 +b\, \delta'(x)$ are
\begin{equation}\label{9}
\left(\begin{array}{c} \psi(+0) \\ \psi'(+0) \end{array}  \right) = 
 \left(\begin{array}{cc} \frac{2-b}{2+b} & 0 \\ 0 & \frac{2+b}{2-b}  \end{array}  \right) 
 \left(\begin{array}{c}  \psi(-0)  \\ \psi'(-0)  \end{array}  \right)
\end{equation}
for the local $\delta'$  \cite{GMMN} and
\begin{equation}\label{10}
\left(\begin{array}{c} \psi(+0) \\ \psi'(+0) \end{array}  \right) = 
\left(\begin{array}{cc} 1 & \beta \\ 0 & 1 \end{array}  \right)
\left(\begin{array}{c}  \psi(-0)  \\ \psi'(-0)  \end{array}  \right)\,.
\end{equation}
for the nonlocal $\delta'$, with the coupling constant $\beta$ arising from the renormalisation procedure required to determine the appropriate self-adjoint operator (see \cite{AK,FGGN}).

Note that, while the local $\delta'$-potential is compatible with the $\delta$-potential in the sense that a Hamiltonian like $H_0 + \delta(x) + \delta'(x)$ (the perturbations may also have real coefficients) admits a self-adjoint determination, this is not possible with the nonlocal $\delta'$-perturbation. The reason is the presence of $\beta$ in the second entry of the first row in the square matrix in \eqref{10}. 

In the present paper, we introduce a one-dimensional Hamiltonian, in which  $H_0=-d^2/dx^2$ is now decorated with two nonlocal $\delta'$-perturbations supported at two centres located at $x_0>0$ and $-x_0$. 

Before moving to a brief description of the sections of this article, we wish to provide our main motivation for the latter. As is well known, point interactions were historically introduced in Quantum Mechanics in order to replace sharply peaked potentials, so that the related Hamiltonians may become solvable models. As a consequence, it would be reasonable to expect that point interactions should {\it always} behave like the short range potentials they are supposed to mimic. As fully attested by the classical Quantum Chemistry textbook example of H$^+_2$ smoothly approaching He$^+$ in the limit $R \to 0^+$ (see \cite{Qc,BB,KH2+,KDW}), two three-dimensional interactions with nonzero range coalesce smoothly as the distance between their centres vanishes. However, as was rigorously proved in some previous papers by our group \cite{AFR,AFR1,AFR3}, a similar phenomenon does not occur for two three-dimensional $\delta$-interactions (see also a very recent contribution to this topic \cite{FST}).The same pathological behaviour is exhibited by two-dimensional $\delta$-interactions \cite{FPR}, as well as by one-dimensional $\delta$-interactions perturbing  the aforementioned Salpeter Hamiltonian $\sqrt{-d^2/dx^2 + m^2}$ \cite{AFR2}. In all these papers it was shown that, by making the coupling parameter suitably dependent on $x_0>0$, these regularised point interactions merge smoothly, exactly like short range interactions. In this article we wish to investigate the behaviour of two nonlocal $\delta'$-interactions in this regard, given that, to the best of our knowledge, the issue has not been dealt with in the existing literature.

The paper is organised as follows: after determining in a rigorous way the self-adjoint Hamiltonian making sense of the merely heuristic expression \eqref{3} in Section 2, we show that, in addition to its absolutely continuous spectrum $[0,+\infty)$, its discrete spectrum consists of two eigenvalues (energy levels) which are functions of the two parameters appearing in the resolvent of such a Hamiltonian, namely $x_0>0$ and $\beta$, the coupling parameter arising as a result of the required renormalisation procedure. While we analyse in detail the dependence of both eigenvalues on $x_0>0$ for any fixed value of $\beta$ in Section 3, we investigate their behaviour as functions of $\beta$ for any fixed value of $x_0>0$ in Section 4. Section 5 is devoted to the study of the resonances of the model. In Section 6 we rigorously prove that, as $x_0 \to 0^+$, the resolvent of the self-adjoint Hamiltonian converges in norm to that of the self-adjoint operator making sense of the merely heuristic expression $-\frac {d^2}{dx^2}-  2 \lambda |\delta'_0 \rangle\langle\delta'_0 |$, which implies that, despite their extremely singular nature, two nonlocal $\delta'$-interactions coalesce smoothly as the distance between their centres is shrunk to zero. Finally, in Section 7, in addition to our concluding remarks, we discuss the remarkable result achieved in Section 6 in relation to our previous works on singular double wells consisting of $\delta$-interactions in $d=1,2,3$ dimensions.

\section{On the rigorous definition of the Hamiltonian $H(\lambda,x_0)$}

Our first objective is to obtain by means of a rigorous procedure a self-adjoint determination of the merely heuristic Hamiltonian $H(\lambda,x_0)$ written below. This procedure is known as renormalisation of the coupling constant. The novelty of this Hamiltonian is that its interaction term is given by two nonlocal $\delta'$-interactions with equal strength and symmetrically located with respect to the origin, that is to say:
\begin{equation}\label{11}
H(\lambda,x_0)=-\frac {d^2}{dx^2}-\lambda \left[|\delta' _{-x_0}\rangle\langle\delta' _{-x_0}|+|\delta' _{x_0}\rangle\langle\delta' _{x_0}|\right]\,,
\end{equation}
for any $x_0>0 $.   It is noteworthy that such an extremely singular point interaction exists only for one-dimensional systems, as can be seen in \cite{AGHH,AK,ADK}. It is worth pointing out that the semiclassical limit of the self-adjoint Hamiltonian making sense of the merely heuristic expression $-\frac {d^2}{dx^2}-\lambda |\delta' \rangle\langle\delta' |$ has recently been investigated in \cite{CAFEPO}.

In previous papers \cite{AFR2,AFR3,FPR,FR}, our group has investigated models consisting of symmetric double wells with $\delta$-interactions in dimensions $d=1,2,3$.  he one-dimensional model with such a double well was also investigated in \cite{KS,S,SCorr}.  Throughout the present paper, we shall carry out our calculations in $p$-space, the momentum space, instead of the more widely used  $x$-space. 

As was seen in the previous section, for the nonlocal $\delta'$-interaction centred at $x_0>0$, by setting $g=f \in {\mathcal S}(\mathbb R)$ in \eqref{5}, we have:
\begin{equation}\label{12}
\left(f,|\delta'_{x_0}\rangle\langle\delta'_{x_0}|f\right)=\left(f',|\delta_{x_0}\rangle\langle\delta_{x_0}|f'\right)=\left(f',\delta_{x_0}\, f' \right)=\int_{-\infty}^{\infty}\delta(x-x_0) \left[f'(x)\right]^2 \, dx=\, \left[f'(x_0)\right]^2\,.
\end{equation}
In $p$-space, \eqref{12} is written as
\begin{equation}\label{13}
\frac 1{2\pi} \left(p\hat{f},e^{ix_0\,p}\right) \left(e^{ix_0\,p},\, p\hat{f}\right)=\frac 1{2\pi} \left| \left(e^{ix_0\,p},\, p\hat{f}\right) \right|^2\,,
\end{equation}
where $\hat{f}(p)$ denotes the Fourier transform of $f(x)$.

Then, for any $x_0>0$, we have the following identities:
\begin{eqnarray}\label{14}
&& \left(f,|\delta'_{-x_0}\rangle\langle\delta'_{-x_0}|f\right)+\left(f,|\delta'_{x_0}\rangle\langle\delta'_{x_0}|f\right) = \left(f',\delta_{-x_0}\, f' \right)+\left(f',\delta_{x_0}\, f' \right)  \nonumber \\ [1.2ex] 
&& \qquad  = \int_{-\infty}^{\infty}\delta(x+x_0) \left[f'(x)\right]^2  dx+\int_{-\infty}^{\infty}\delta(x-x_0) \left[f'(x)\right]^2  dx   = \left[f'(-x_0)\right]^2+\left[f'(x_0)\right]^2 \,.
\end{eqnarray}
In terms of momenta, the latter expression is equal to
\begin{equation}\label{15}
\frac 1{2\pi} \left[ \left| \left(e^{ix_0\,p},\, p\hat{f}\right) \right|^2 + \left| \left(e^{-ix_0\,p},\, p\hat{f}\right) \right|^2 \right]\,,
\end{equation}
which shows that $|\delta'_{-x_0}\rangle\langle\delta'_{-x_0}|+|\delta'_{x_0}\rangle\langle\delta'_{x_0}|$ is a rank two perturbation \cite{AK}. 

As attested by \cite{KLA,FA,FGNR}, the operator
\begin{equation}\label{16}
B_{x_0;E}: = \left[-\frac {d^2}{dx^2}\, + |E|\, \right]^{-\frac 12} V(\cdot) \left[-\frac {d^2}{dx^2}\, + |E|\, \right]^{-\frac 12}
\end{equation}
is isospectral to the more commonly used Birman-Schwinger operator
\begin{equation}
\text{sgn} (V)\,|V|^{\frac 12}\,\left[-\frac {d^2}{dx^2}\, + |E|\, \right]^{-1}\,|V|^{\frac 12}.
\end{equation}

Adopting the technique used in references \cite{AFR2,AFR3,FPR,FR}  and taking account once again of the fact that $(f,\delta')(\delta',g)=(f',\delta)(\delta,g')=(f',\delta\, g')$ (see \cite{GLP,AFR4}), we can write the two-dimensional integral kernel of the integral operator in equation \eqref{16} with potential $V:= |\delta'_{-x_0}\rangle\langle\delta'_{-x_0}|+|\delta'_{x_0}\rangle\langle\delta'_{x_0}|$ in momentum space as
\begin{eqnarray}\label{17}
B_{x_0;E}(p,p') \!\!&\!\!=\!\! &\!\!   \frac 1{\pi}\, \frac {p}{\left(p^2+|E|\right)^{1/2}}\cos \left(x_0(p-p')\,\right) \frac {p'}{\left(p'^2+|E|\right)^{1/2}}
\nonumber \\ [0.5ex]  
\!\!&\!\!=\!\! &\!\! 
 \frac 1{\pi}\, \frac {p}{\left(p^2+|E|\right)^{1/2}}\,\left[ \cos x_0 p\, \cos x_0 p'\, + \sin x_0 p \,  \sin x_0 p'\, \right] \frac {p'}{\left(p'^2+|E|\right)^{1/2}}\, , 
\end{eqnarray}
since the Fourier transform of $\frac {df}{dx}$ is given by $ip\hat{f}$.

The simplicity of the latter expression shows rather explicitly why in this context it is more convenient to use the operator $B_{x_0;E}$ instead of the Birman-Schwinger operator.
It is interesting to compare the above kernel to its counterpart when the interaction term is of the form $\delta(x-x_0) + \delta(x+x_0)$. In momentum space the latter kernel, $b_{x_0;E}(p,p')$, is given by
\begin{equation}\label{18}
b_{x_0;E}(p,p')=\frac 1{\pi}\, \frac 1{\left(p^2+|E|\right)^{1/2}}\, \cos\left(x_0(p-p')\,\right) \frac 1{\left(p'^2+|E|\right)^{1/2}}\,.
\end{equation}

As rigorously shown in \cite{FR}, $b_{x_0;E}(p,p')$ is the kernel of a trace class positive integral operator, which is actually a rank two operator, acting on $L^2(\mathbb R)$ with trace norm equal to
\begin{equation}\label{19}
\| b_{x_0;E}\|_{T_1}=\frac 1{\pi} \int_{-\infty}^{\infty}\, \frac 1{p^2+|E|}\, dp=\, \frac 2{\pi} \int_{0}^{\infty}\,  \frac 1{p^2+|E|}\, dp=\, \frac 1{|E|^{1/2}}\,.
\end{equation}

This result has an interesting conclusion \cite{AGHH,AK,RSI,RSII}, which is that the {\it quadratic form} domain of a self-adjoint determination of the Hamiltonian\footnote{The {\it operator} domain of this self-adjoint determination is the space of functions on a Sobolev space verifying \eqref{8} at each of the points $x_0$ and $-x_0$.} $H=-d^2/dx^2 + \delta(x-x_0) + \delta(x+x_0)$ coincides with that of the unperturbed $H_0=-d^2/dx^2$ since the finiteness of the trace implies that the renowned KLMN theorem is immediately applicable (see \cite{RSII,DONSIMON}). 

On the other hand, the operator with kernel $B_{x_0;E}(p,p')$ in \eqref{17} is not trace class. Nevertheless, let us consider the following positive operator
\begin{eqnarray}\label{20}
\left(p^2+|E|\right)^{-\frac 12}B_{x_0;E}\left(p^2+|E|\right)^{-\frac 12}  =  \left(p^2+|E|\right)^{-1}\left[ |\delta'_{-x_0}\rangle\langle\delta'_{-x_0}|+
|\delta'_{x_0}\rangle\langle\delta'_{x_0}| \right] \left(p^2+|E|\right)^{-1}\,,
\end{eqnarray}
having a form similar to \eqref{16}. This operator has the following kernel in momentum space

\begin{equation}\label{21}
\frac 1{\pi}\, \frac {p}{p^2+|E|}\, \cos\left(x_0(p-p')\,\right) \frac {p'}{p'^2+|E|}\,.
\end{equation}
The latter positive operator has a finite trace given by
\begin{eqnarray}\label{22}
&&\hskip-1.9cm   \frac 1{\pi}\,\int_{-\infty}^{\infty}\, \frac {p^2\, dp}{\left(p^2+|E|\right)^2}  = 
\frac 2{\pi}\,\int_0^{\infty}\, \frac {p^2\, dp}{\left(p^2+|E|\right)^2} =  \frac 2{\pi}\left[ \int_0^{\infty}\, \frac{dp}{p^2+|E|} - \, |E|\int_0^{\infty}\, \frac{dp}{\left(p^2+|E|\right)^2}\right]  \nonumber \\ [1.ex] 
&&   = 
   \frac 2{\pi}\left[ \frac {\pi}{2|E|^{1/2}}\, - \, |E|\int_0^{\infty}\, \frac 1{\left(p^2+|E|\right)^2}\, dp\right]    =
    \frac 2{\pi}\left[ \frac {\pi}{2|E|^{1/2}}\, - \,\frac {\pi}{4|E|^{1/2}} \right]=\frac 1{2|E|^{1/2}}\,.
\end{eqnarray}
This shows that $|\delta'_{-x_0}\rangle\langle\delta'_{-x_0}|+|\delta'_{x_0}\rangle\langle\delta'_{x_0}|$, while not in $\mathcal H_{-1}$, is in $\mathcal H_{-2}$ since for any $f \in \, \mathcal H_{2}$:
\begin{eqnarray}
&&\hskip-0.7cm  \left(f,\,\left[ |\delta'_{-x_0}\rangle\langle\delta'_{-x_0}|+|\delta'_{x_0}\rangle\langle\delta'_{x_0}| \right] \, f\,\right)= \nonumber  \\
&&\hskip-0.7cm  =\left(\left[-\frac {d^2}{dx^2}\, + |E|\, \right]f,\,\left[-\frac {d^2}{dx^2}\, + |E|\, \right]^{-1}\left[ |\delta'_{-x_0}\rangle\langle\delta'_{-x_0}|+|\delta'_{x_0}\rangle\langle\delta'_{x_0}| \right] \left[-\frac {d^2}{dx^2}\, + |E|\, \right]^{-1}\, \left[-\frac {d^2}{dx^2}\, + |E|\, \right]f\,\right) \nonumber  \\
&&\hskip-0.7cm  =\left(\left(p^2+|E|\right)\,\hat f, \left(p^2+|E|\right)^{-\frac 12}B_{x_0;E}\left(p^2+|E|\right)^{-\frac 12} \,\left(p^2+|E|\right)\hat f\right)\leq\frac 1{2|E|^{1/2}}\,\|\left(p^2+|E|\right)\,\hat f\|_2^2 .
\end{eqnarray}

This property has a consequence: as has been explained in detail in \cite{AK} (Lemma 1.2.2), rank one perturbations defined by vectors that are in $\mathcal H_{-2}$ but not in $\mathcal H_{-1}$ are not form bounded with respect to the self-adjoint operator $-\frac {d^2}{dx^2}$ defined on the first Sobolev space $\mathcal H_1$. In accordance with \cite{AK}, this implies that, in order to achieve a self-adjoint determination of the merely heuristic Hamiltonian in \eqref{11}, either the theory of self-adjoint extensions of symmetric operators or the renormalisation of the coupling constant is required.  While the former might be the favourite choice of mathematicians, we have opted to use the latter in order to highlight the analogy between the model studied here and those with a singular double well consisting of two identical attractive $\delta$-interactions as a perturbation of either the semirelativistic Salpeter Hamiltonian in one dimension or the negative Laplacian in two or three dimensions, that is to say the necessity of fixing the ultraviolet divergences (short distances or, equivalently, large momenta) arising because of the point interactions, as attested by articles such as \cite{Cav} in addition to the aforementioned articles. 

It is worth recalling that, as was rigorously demonstrated in \cite{FR,EX}, the Hamiltonian with a nonlocal $\delta'$-interaction is the norm resolvent limit of a net of Hamiltonians with a suitable triple of $\delta$ interactions. Therefore, as fully shown in \cite{EX}, the Hamiltonian with a nonlocal $\delta'$-interaction can be approximated by Hamiltonians with the interaction term given by the sum of three short range potentials with shrinking supports. The latter fact implies that the approximation of our Hamiltonian with a singular double well with a pair of identical nonlocal $\delta'$-interactions by means of Hamiltonians with short range potentials, although feasible in principle, would not be very practical from the operational point of view since one should use six short range potentials with shrinking supports.

Then, we consider the following Hamiltonian in which we have introduced a cutoff for large values of the momentum:
\begin{eqnarray}\label{23}
H(k,\lambda(k),x_0)  =   p^2-\lambda(k)\,  p\, \chi_{|p|<k}(p)  \left[|e^{-ix_0p}\rangle\langle e^{-ix_0p}|+|e^{ix_0p}\rangle\langle e^{ix_0p}|\right]p\, \chi_{|p|<k}(p)\,.
\end{eqnarray}
Here, the function $\chi_{|p|<k}(p) $ is the characteristic function of the set of momenta with magnitude less than
the cutoff set at $k$. Observe that the constant $\lambda$ in \eqref{1} is now a function, $\lambda(k)$, to be determined from the value of the maximum momentum. After the removal of this ultraviolet divergence, it results that the Hamiltonian $H(k,\lambda(k),x_0)$ is a perfectly defined self-adjoint operator. 

Next, we go back to the operator $B_{x_0;E}$ with integral kernel \eqref{17} and apply on it the previous ultraviolet cutoff. Thus, we obtain an operator denoted by  $B^k_{x_0;E}$, where $k$ is defined as before. Following the procedure in \cite{AFR2}, we can determine the following resolvent operator:
\begin{eqnarray}\label{24}
\left[I-\lambda(k)B_{x_0;E}^{k}\right]^{-1} \!\!&\!\!=\!\! &\!\!    I+ \frac 1{\frac {\pi}{\lambda(k)}-\, \int_{-k}^{k}\frac {p^2\,\sin^2x_0 p}{p^2+|E|} dp}\, \left| \frac {\chi_{|p|<k}\,p\, \sin x_0 p}{\left(p^2+|E|\right)^{1/2}} \right> \left< \frac {\chi_{|p|<k}\,p\, \sin x_0 p}{\left(p^2+|E|\right)^{1/2}} \right| \nonumber \\ [1.ex]   
\!\!&\!\!  \!\! &\!\! 
+ \frac 1{\frac {\pi}{\lambda(k)}-\, \int_{-k}^{k}\frac {p^2\,\cos^2x_0 p}{p^2+|E|} dp}\, \left| \frac {\chi_{|p|<k}\,p\, \cos x_0 p}{\left(p^2+|E|\right)^{1/2}} \right> \left< \frac {\chi_{|p|<k}\,p\, \cos x_0 p}{\left(p^2+|E|\right)^{1/2}} \right|   \nonumber \\ [1.ex]  
\!\!&\!\!=\!\! &\!\! 
 I+ \frac 1{\frac {\pi}{\lambda(k)}-2\, \int_{0}^{k}\frac {p^2\,\sin^2x_0 p}{p^2+|E|} dp}\, \left| \frac {\chi_{|p|<k}\,p\, \sin x_0 p}{\left(p^2+|E|\right)^{1/2}} \right> \left< \frac {\chi_{|p|<k}\,p\, \sin x_0 p}{\left(p^2+|E|\right)^{1/2}} \right|   \nonumber \\ [1.ex]  
 \!\!&\!\! \!\! &\!\! 
 + \frac 1{\frac {\pi}{\lambda(k)}-\,2 \int_{0}^{k}\frac {p^2\,\cos^2x_0 p}{p^2+|E|} dp}\, \left| \frac {\chi_{|p|<k}\,p\, \cos x_0 p}{\left(p^2+|E|\right)^{1/2}} \right> \left< \frac {\chi_{|p|<k}\,p\, \cos x_0 p}{\left(p^2+|E|\right)^{1/2}} \right|  \,.
\end{eqnarray}

Observe that in the denominators of the coefficients in \eqref{24}, there are some integrals that should be evaluated before studying their limit as $k \to \infty$. The former integral is
\begin{eqnarray}\label{25}
2\int_{0}^{k}\frac {p^2\,\sin^2x_0 p}{p^2+|E|} dp  
\!\!&\!\!=\!\! &\!\!  
2 \int_{0}^{k}\sin^2x_0 p\, dp-2|E|\,\int_{0}^{k}\frac {\sin^2x_0 p}{p^2+|E|} dp \nonumber \\ [1.2ex]   
\!\!&\!\!=\!\! &\!\! 
 \int_{0}^{k}\left[1-\cos 2x_0 p\right]\, dp\,- |E|\,\int_{0}^{k}\frac {1-\cos 2x_0 p}{p^2+|E|} dp \nonumber \\ [1.2ex]   \!\!&\!\!=\!\! &\!\! 
    k\left(1-\frac {\sin 2x_0 k}{2x_0k}\right)\,-\, |E|^{1/2}\, \tan^{-1}\left(\frac {k}{|E|^{1/2}}\right)+ |E|\,\int_{0}^{k}\frac {\cos 2x_0 p}{p^2+|E|} dp\,.
\end{eqnarray}
The latter is
\begin{eqnarray}\label{26}
2\int_{0}^{k}\frac {p^2\,\cos^2x_0 p}{p^2+|E|} dp  \!\!&\!\!=\!\! &\!\! 2 \int_{0}^{k}\cos^2x_0 p\, dp-2|E|\,\int_{0}^{k}\frac {\cos^2x_0 p}{p^2+|E|} dp    \nonumber \\ [1.2ex]   
\!\!&\!\!=\!\! &\!\! 
  \int_{0}^{k}\left[1+\cos 2x_0 p\right]\, dp\,- |E|\,\int_{0}^{k}\frac {1+\cos 2x_0 p}{p^2+|E|} dp \nonumber \\ [1.2ex]   \!\!&\!\!=\!\! &\!\! 
     k\left(1+\frac {\sin 2x_0 k}{2x_0k}\right)\,-\, |E|^{1/2}\, \tan^{-1}\left(\frac {k}{|E|^{1/2}}\right)- |E|\,\int_{0}^{k}\frac {\cos 2x_0 p}{p^2+|E|} dp\,.
\end{eqnarray}

To understand the notation used in \eqref{24}, let us assume that $f(x) \in L^2(\mathbb R)$. As is well known, the expression $|f\rangle\langle f|$ defines a rank one operator on $L^2(\mathbb R)$, for if $g$ is arbitrary in $L^2(\mathbb R)$, then, the action of $|f\rangle\langle f|$ on $g$ is defined as $(f,g)\,|f\rangle$, where $(f,g)$ is the scalar product of $f$ with $g$, so that $|f\rangle\langle f|$ is the orthogonal projection of $L^2(\mathbb R)$ on the one-dimensional subspace spanned by the function $f$.

Now, it is time to fix the function $\lambda(k)$. Following \cite{AGHH,AK}, we fix for some $\beta \ne 0$,
\begin{equation}\label{27}
\frac {\pi}{\lambda(k)}=k+\frac {\pi}{\beta}\,,
\end{equation}
so that
\begin{equation}\label{28}
\lambda(k)=\frac {\beta\, \pi}{\beta k+\pi}\,.
\end{equation}

After \eqref{27}--\eqref{28} and taking into account \eqref{25}--\eqref{26}, we may find the limits as $k \to \infty$ of the denominators of the coefficients in \eqref{24}. For the first denominator, we have
\begin{equation}\label{29}
\frac {\pi}{\lambda(k)}-\,2 \int_{0}^{k}\frac {p^2\, \sin^2x_0 p}{p^2+|E|} dp \to   \frac {\pi}{\beta} + \frac {\pi\, |E|^{1/2}}{2}\, \left(1-e^{-2x_0|E|^{1/2}} \right)\,.
\end{equation}
For the second,
\begin{equation}\label{30}
\frac {\pi}{\lambda(k)}-\,2 \int_{0}^{k}\frac {p^2\,\cos^2x_0 p}{p^2+|E|} dp    \to    \frac {\pi}{\beta} + \frac {\pi\, |E|^{1/2}}{2}\, \left(1+e^{-2x_0|E|^{1/2}} \right)\,,
\end{equation}
where the improper integrals that appear after the limit have been evaluated in \cite{FR}. Consequently, in the limit $k \to \infty$, we have that
\begin{eqnarray}\label{31}
&&\left[I-\lambda(k)B_{x_0;E}^{k}\right]^{-1} \to    I + \frac 1{\frac {\pi}{\beta} + \frac {\pi\, |E|^{1/2}}{2}\, \left(1-e^{-2x_0|E|^{1/2}} \right)}\, \left| \frac {p\, \sin x_0 p}{\left(p^2+|E|\right)^{1/2}} \right> \left< \frac {p\sin x_0 p}{\left(p^2+|E|\right)^{1/2}} \right|   \nonumber 
\\ [1ex]    
 && \hskip3.5cm
 +\frac 1{\frac {\pi}{\beta} + \frac {\pi\, |E|^{1/2}}{2}\, \left(1+e^{-2x_0|E|^{1/2}} \right)} \left| \frac {p\cos x_0 p}{\left(p^2+|E|\right)^{1/2}} \right> \left< \frac {p\cos x_0 p}{\left(p^2+|E|\right)^{1/2}} \right| \,.
\end{eqnarray}
The latter expression does not define a bounded operator on $L^2(\mathbb R)$ since the functions inside the rank one operators are manifestly far from being square summable. However, $\left[I-\lambda(k)B_{x_0;E}^{k}\right]^{-1}$ {\it per se} is not physically relevant.  Nevertheless, we may exploit Tiktopoulos' formula (see \cite{RSI,DONSIMON}), that is to say
\begin{eqnarray}\label{32}
\left(H_0-V+|E|\right)^{-1}     =\left(H_0+|E|\right)^{-1/2}\left[I-\left(H_0+|E|\right)^{-1/2}\,V\, \left(H_0+|E|\right)^{-1/2}\right]^{-1}\left(H_0+|E|\right)^{-1/2}, 
\end{eqnarray}
valid for any positive $H_0\ge 0$ and potential $V\ge 0$, in order to write the resolvent of $H(k,\lambda(k),x_0)$ and then perform its limit as $ k \to \infty$:
\begin{eqnarray}\label{33}
&& \hskip-2cm\left[H(k,\lambda(k),x_0)+|E|\right]^{-1} = \left(p^2+|E|\right)^{-1/2}\,\left[I-\lambda(k)B_{x_0;E}^{k}\right]^{-1}\, \left(p^2+|E|\right)^{-1/2} \;\;\;\;  \nonumber \\ [1.2ex]  
&& \quad \to   \left(p^2+|E|\right)^{-1}+\frac 1{\pi\, \left[\frac 1{\beta} + \frac {|E|^{1/2}}{2}\, \left(1-e^{-2x_0|E|^{1/2}} \right) \right]}\,  \left| \frac {p\, \sin x_0 p}{p^2+|E|} \right> \left< \frac {p\, \sin x_0 p}{p^2+|E|} \right| \;\;\;\;  \nonumber \\ [1.2ex]  
 &&\qquad \quad
 + \frac 1{\pi\, \left[\frac 1{\beta} + \frac {|E|^{1/2}}{2}\, \left(1+e^{-2x_0|E|^{1/2}} \right) \right]}\,  \left| \frac {p\, \cos x_0 p}{p^2+|E|} \right> \left< \frac {p\, \cos x_0 p}{p^2+|E|} \right|     =: R(\beta,x_0,|E|)\,,
\end{eqnarray}
where the last identity defines the operator valued function $R(\beta,x_0,|E|)$. The functions that determine the rank one operators are square integrable since the squares of both functions are bounded by the integrand in:
\begin{eqnarray}\label{34}
\int_{0}^{+\infty} \frac {p^2\, dp}{\left(p^2+|E|\right)^2} = \int_{0}^{+\infty} \frac 1{p^2+|E|}\, dp\,-2|E|\, \int_{0}^{+\infty}  \frac{\, dp}{\left(p^2+|E|\right)^2}   =  \frac {\pi}{2|E|^{1/2}}\, - \,\frac {\pi}{4|E|^{1/2}}  = \frac {\pi}{4|E|^{1/2}}
\end{eqnarray}

By proceeding essentially along the lines of the proof of Lemma 3.1. in \cite{AFR5}, we may show that the limit in \eqref{33} is given in the norm of trace class operators on $L^2(\mathbb R)$, so that
\begin{equation}\label{35}
\| R(\beta,x_0,|E|)\, - \left[H(k,\lambda(k),x_0)+|E|\right]^{-1}\, \|_{T_1} \to 0\,, \quad k \to \infty\,
\end{equation}

Finally, we should prove that the operator $R(\beta,x_0,|E|)$ is indeed the resolvent of a self-adjoint operator. However, such a proof will be omitted here since it would be essentially identical to those of Theorem~1.1.1 Ch. II.1 in \cite{AGHH}, Theorem 2.2 in \cite{FI2}, and Theorem 2.1 in \cite{AFR2}.

\begin{figure}[htb]
\begin{center}
\includegraphics[width=0.4\textwidth]{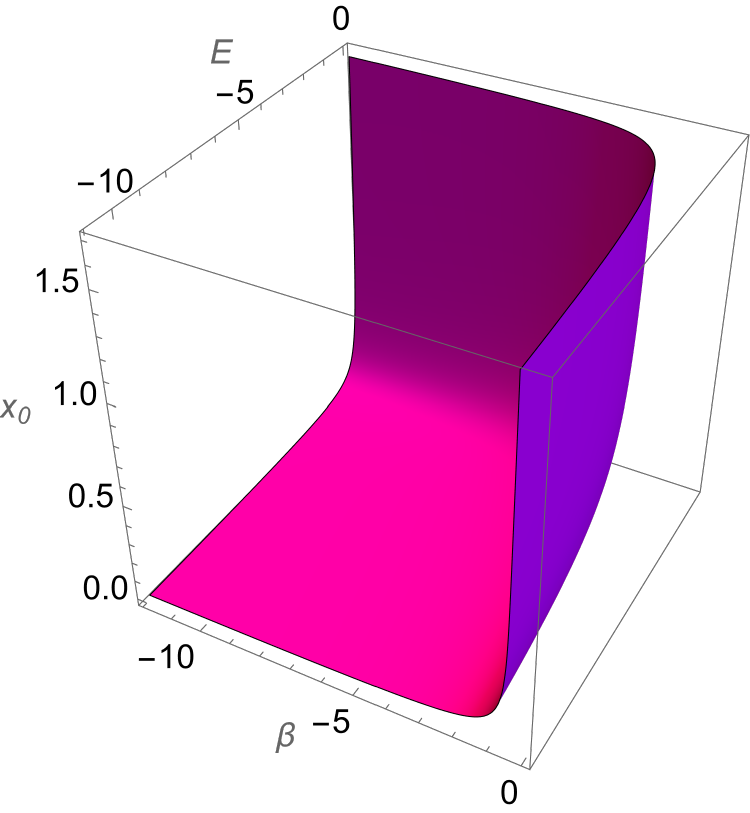}
\qquad\qquad
\includegraphics[width=0.4\textwidth]{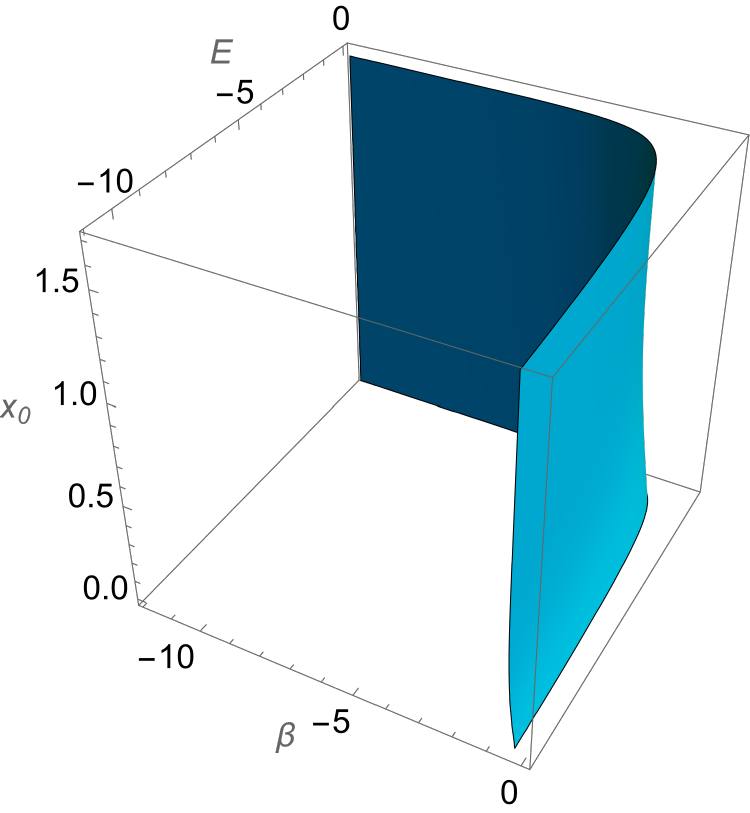}
\caption{3D plots of the ground state energy $E_0(x_0,\beta)$ and the excited state energy $E_1(x_0,\beta)$ as functions of $x_0$ and $\beta$. }
\label{figure1}
\end{center}
\end{figure}

As fully attested by \eqref{33}, $R(\beta,x_0,|E|)$ is a rank two perturbation of the free resolvent, so that it is straightforward to infer that the operator domain of the limiting operator in $p$-space consists of all the vectors in the operator domain of the free Hamiltonian and the two-dimensional subspace spanned by the vectors
$$
\hat f_0(p;x_0,E_0(x_0,\beta))=\,  \frac {p\, \sin x_0 p}{p^2+|E_0(x_0,\beta)|}, \qquad 
\hat f_1(p;x_0,E_1(x_0,\beta))=\,  \frac {p\, \cos x_0 p}{p^2+|E_1(x_0,\beta)|},
$$ 
which implies that the operator domain in $x$-space consists of all the vectors in $\mathcal H_{2}$, the second Sobolev space, and the two-dimensional subspace spanned by the vectors
\begin{eqnarray}
&& f_0(x;x_0,E_0(x_0,\beta))=\, \frac 1{2|E_0(x_0,\beta)|^{1/2}}\,\frac {d}{dx}\, \left(e^{-|E_0(x_0,\beta)|^{1/2}|x-x_0|}-e^{-|E_0(x_0,\beta)|^{1/2}|x+x_0|}\right),
\\ [1ex]
&& f_1(x;x_0,E_1(x_0,\beta))=\, \frac 1{2|E_1(x_0,\beta)|^{1/2}}\,\frac {d}{dx}\, \left(e^{-|E_1(x_0,\beta)|^{1/2}|x-x_0|}+e^{-|E_1(x_0,\beta)|^{1/2}|x+x_0|}\right),
\end{eqnarray}
as follows from (2.10a) in \cite{FR}. Here $E_0(x_0,\beta)$ (respectively $E_1(x_0,\beta)$) is the zero of the denominator of the second (resp. third) term in \eqref{33}, so that it is nothing else but the ground state (resp. excited state) eigenenergy. The eigenvalues  $E_0(x_0,\beta)$ and $E_1(x_0,\beta)$, depicted as functions of both parameters in Figure~\ref{figure1}, will be studied in detail in the next two sections.

\smallskip

Let us summarise the results of the present Section as follows:

\noindent
{\bf Theorem 1.} {\it
The rigorous Hamiltonian making sense of the merely heuristic expression
\begin{equation}\label{36}
H(\lambda,x_0)=-\frac {d^2}{dx^2}-\lambda \left[|\delta'_{-x_0}\rangle\langle\delta'_{-x_0}|+|\delta'_{x_0}\rangle\langle\delta'_{x_0}|\right], \quad x_0>0 ,
\end{equation}
is the self-adjoint operator $H_{sa}(\beta,x_0)$  whose resolvent is given by $R(\beta,x_0,|E|)$, defined in \eqref{33}, for any $E<0, \beta \neq 0, x_0>0$.
The latter is the limit, as $k \to +\infty$, in norm convergence of the resolvents of the Hamiltonians 
$H(k,\lambda(k),x_0)$ with the ultraviolet momentum cutoff defined by \eqref{27} or, equivalently, \eqref{28}. Furthermore, $H_{sa}(\beta,x_0)$, regarded as a function of $\beta$, is an analytic family in the sense of Kato.
}

\section{On the eigenvalues of $H_{sa}(\beta,x_0)$ as functions of $x_0>0$}

In this section we shall assume that $x_0>0$ and the coupling parameter $\beta<0$ is fixed. The eigenvalues of $H_{sa}(\beta,x_0)$ are determined by the poles along the negative semiaxis $E<0$ of its resolvent $R(\beta,x_0,|E|)$. Thus, throughout the present Section, we shall investigate in detail the two equations determining the two eigenvalues created by the the singular double well, namely the unique solution, for any {\it fixed} $\beta < 0, x_0>0$ and $E<0$, of
\begin{equation}\label{37}
\frac 1{\beta} + \frac {|E|^{1/2}}{2}\, \left(1-e^{-2x_0|E|^{1/2}} \right)=0\,,
\end{equation}
for the ground state energy and the unique solution of
\begin{equation}\label{38}
\frac 1{\beta} + \frac {|E|^{1/2}}{2}\, \left(1+e^{-2x_0|E|^{1/2}} \right)=0\,, 
\end{equation}
for the energy of the excited state.

Here a brief remark is needed. By reviewing the collection of papers previously published by our group, one may compare equations \eqref{37}--\eqref{38} with some results obtained in \cite{FR}, in particular equations (2.12) and (2.11), in which a Hamiltonian similar to \eqref{3}, with the delta primes replaced by the deltas, was investigated.  Furthermore, it is worth recalling that in \cite{FGNR1} we encountered similar equations for the self-adjoint Hamiltonians acting on  $L^2(\mathbb R^+)$
\begin{equation}\label{39}
\left[-\frac {d^2}{dx^2}\right]_{D}-\lambda \delta (x-x_0)\,, \qquad \left[-\frac {d^2}{dx^2}\right]_{N}-\lambda \delta (x-x_0)
\end{equation}
with $x_0>0$. Here, the subindices $D$ and $N$ stand for Dirichlet and Neumann boundary conditions at the origin, respectively. 

For any fixed value of $\beta$, we wish to obtain the energy as a function of $x_0$. We first note that \eqref{37} can be written as
\begin{equation}\label{40}
x_0(E_0)=-\frac 1{2 |E_0|^{1/2}}\, \ln \left(1- \frac 2{|\beta|\, |E_0|^{1/2}}\right)>0\,,
\end{equation}
which is well defined provided that $E_0<-4/\beta^2$. Analogously, expression \eqref{38} admits the following version:
\begin{equation}\label{41}
x_0(E_1)=-\frac 1{2 |E_1|^{1/2}}\, \ln \left(\frac 2{|\beta|\, |E_1|^{1/2}}-1\right)>0\,,
\end{equation}
provided that $-\frac{4}{\beta^2} < E_1 < -\frac{1}{\beta^2}$. 

\begin{figure}[htb]
\begin{center}
\includegraphics[width=0.4\textwidth]{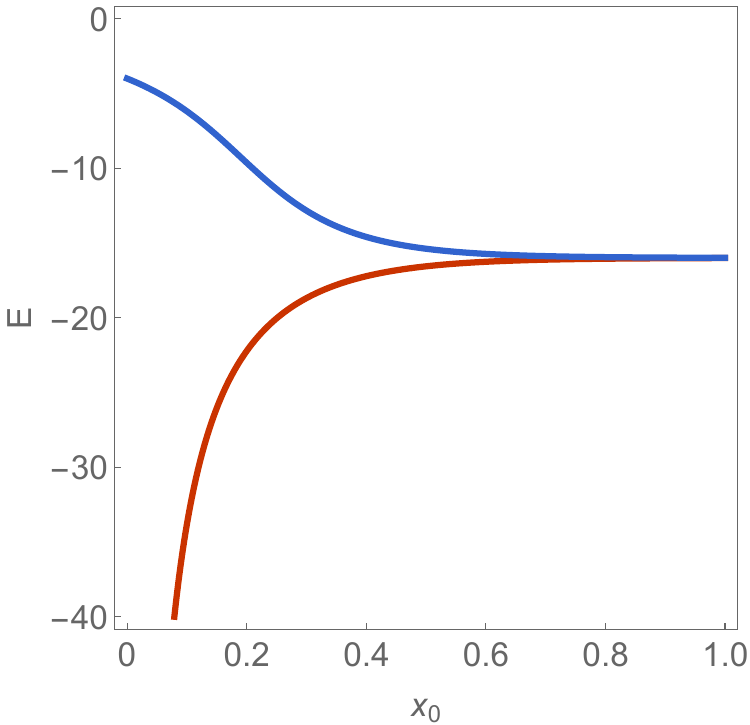}
\qquad\qquad
\includegraphics[width=0.4\textwidth]{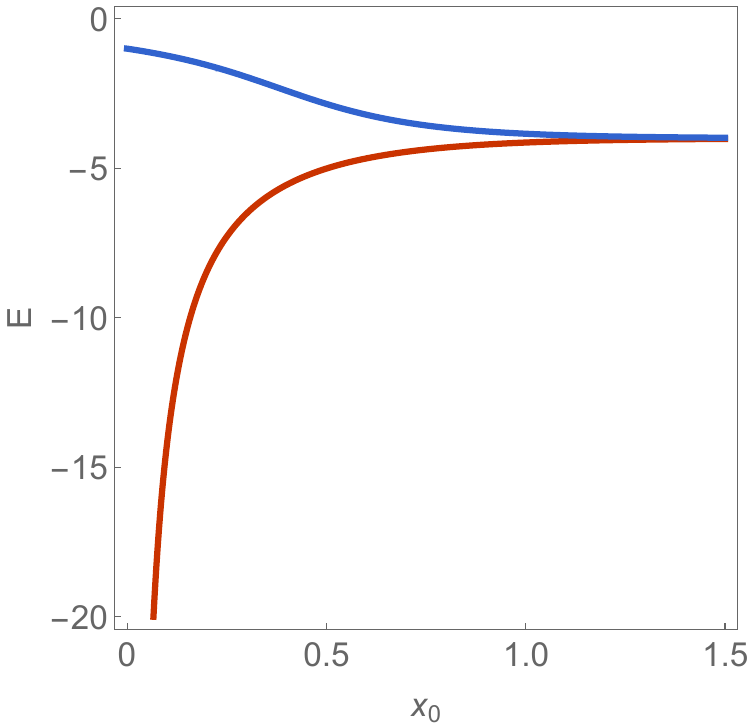}
\caption{Plot of the ground state energy $E_0(x_0)$ (red curve) and the excited state energy $E_1(x_0)$ (blue curve) as functions of $x_0$ for $\beta=-1/2$ (left) and $\beta=-1$ (right). }
\label{figure3}
\end{center}
\end{figure}

\begin{figure}[htb]
\begin{center}
\includegraphics[width=0.4\textwidth]{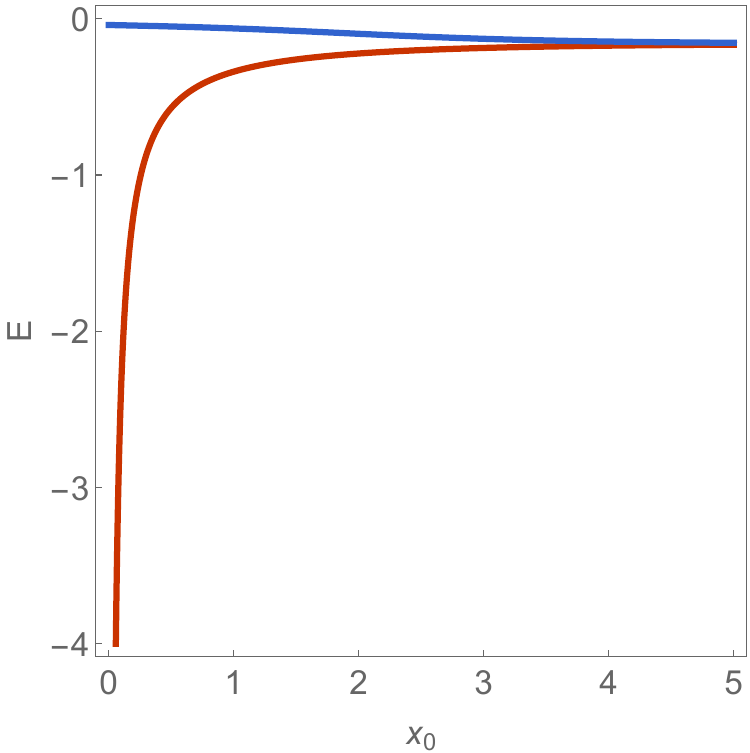}
\quad\qquad
\includegraphics[width=0.4\textwidth]{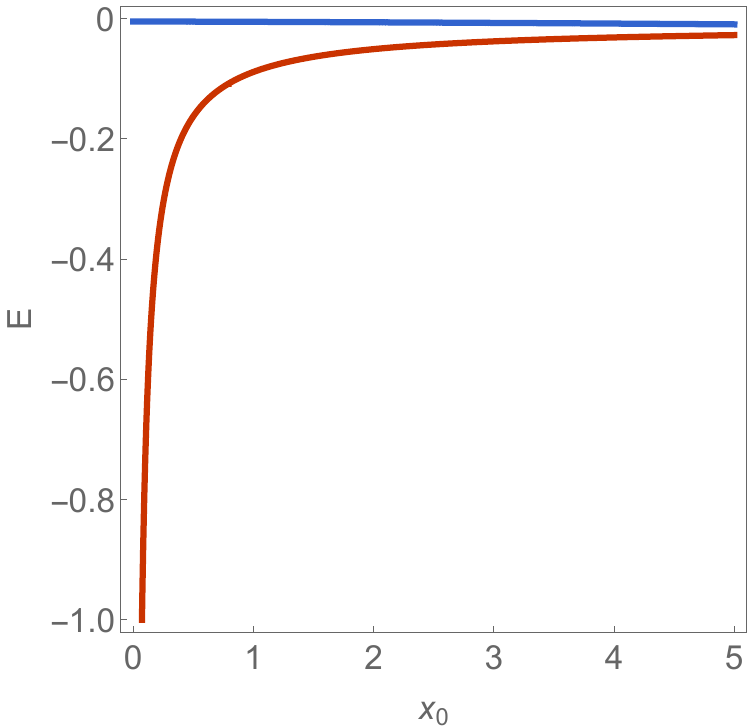}
\caption{Plot of the ground state energy $E_0(x_0)$ (red curve) and the excited state energy $E_1(x_0)$ (blue curve) as functions of $x_0$ for $\beta=-5$ (left) and $\beta=-15$ (right).}
\label{figure5}
\end{center}
\end{figure}

It is easy to check that both $x_0(E_0)$ and $x_0(E_1)$ are strictly monotonic in their arguments, $E_0$ and $E_1$, respectively. Therefore, they are invertible, so that one may find the ground state energy, $E_0(x_0)$, as a function of $x_0>0$ within the range $\left(-\infty,-\frac 4{\beta^2}\right)$ and the first excited state, $E_1(x_0)$, on the interval $\left(-\frac 4{\beta^2},-\frac 1{\beta^2}\right)$. 
These curves, $E_0(x_0)$ and $E_1(x_0)$, are plotted in Figures~\ref{figure3}--\ref{figure5} for various values of $\beta$.

As attested by the plots, $E_1(x_0)$ approaches asymptotically $-\frac 4{\beta^2}$ from above as $x_0 \to +\infty$.  
The ground state energy, $E_0(x_0)$, has the same limit, this time from below. As a consequence, the first excited state energy is practically indistinguishable from the ground state energy for large values of $x_0$, i.e., when both centres are far apart from each other. Therefore, the ionisation energy decreases as the distance between the centres widens and vanishes in the limit $x_0 \to +\infty$.

This spectral feature, that may be called {\it asymptotic degeneracy}, is a feature shared by both the Hamiltonian \eqref{3} and its twin Hamiltonian with the delta primes replaced by deltas, i.e., 
\begin{equation}\label{42}
-\frac {d^2}{dx^2}-\lambda \left[\delta (x+x_0)+\delta (x-x_0) \right]\,,
\end{equation}
studied in detail in \cite{FR}, both eigenvalues of which converge to $-\frac {\lambda^2}4$. The latter is the eigenvalue of the Hamiltonian $-d^2/dx^2- \lambda \delta(x-x_0)$, $\lambda>0$ for any $x_0$ along the real line. In our case, both eigenvalues converge to $-\frac 4{\beta^2}$, which is {\it precisely} the eigenvalue of the self-adjoint determination of the heuristic Hamiltonian $-d^2/dx^2-  \lambda \delta'(x-x_0)$, $\lambda>0$ for any $x_0$ along the real line, $\beta$ being the coupling parameter arising from the renormalisation required to achieve the self-adjoint determination (see \cite{AGHH,AK}). This kind of asymptotic degeneracy also appears in the analysis of the self-adjoint determination of the  semi-relativistic Salpeter Hamiltonian
\begin{equation}\label{43}
\sqrt{p^2+m^2}-\lambda \left[\delta (x+x_0)+\delta (x-x_0) \right]\,, 
\end{equation}
which was analysed in detail in \cite{AFR2}. In this case, the limit value of the eigenvalues as $x_0 \to \infty$ just coincides with the only eigenvalue of the self-adjoint determination of $\sqrt{p^2+m^2}-  \,  \lambda \delta (x-x_0)$,  $\lambda>0$ for any $x_0$ along the real line. The self-adjoint determinations of these two Hamiltonians require an appropriate renormalisation procedure \cite{AFR2,HSW,EGU}. 

The present model shares some spectral properties with other models previously investigated by our group \cite{AFR,AFR1,AFR2,FR,FR1} in the sense that {\it the greater the distance between two impurities is, the less localised the ground state will be}. Also, as stated in \cite{AFR,AFR1,FR1}, the ground state energy behaves similarly even if the free Hamiltonian is given by that of the harmonic oscillator in one, two or three dimensions. A similar phenomenon was observed by Br\"uning et al. \cite{BG} in a study of the ground state energy of the three-dimensional harmonic oscillator with a point perturbation, in particular  with respect to the distance between the location of the bottom of the harmonic potential and that of the point perturbation. It is worth pointing out that this Hamiltonian serves as a model for a three-dimensional quantum dot. 

Once we have mentioned the analogies between the models given by Hamiltonians \eqref{3}, with two delta primes, and \eqref{42}, with two deltas, it is time to stress their differences.  The discrete spectrum of the self-adjoint determination of \eqref{3} consists of two distinct eigenvalues, as follows from \eqref{33} and gets visualised in Figures~\ref{figure1} and \ref{figure3}. As the distance between the centres vanishes, i.e., in the limit $x_0\to 0$, $E_1(x_0)$ converges to $-1/\beta^2$, a value which, for any {\it finite $\beta<0$}, is always below the mimimum of $\sigma_{ac}\left(H_{sa}(\beta,x_0)\,\right)=[0,+\infty)$. Therefore, there is no emergence of this eigenvalue out of the absolutely continuous spectrum of $H_{sa}(\beta,x_0)$ for any {\it finite} negative value of the coupling $\beta$.  On the other hand, the discrete spectrum of \eqref{42} has a bound state and admits a second one, with higher energy, provided that $\lambda x_0>1$, with $\lambda>0$. As rigorously shown in \cite{FPR},  this excited state emerges from the absolutely continuous spectrum of \eqref{42} at $x_0= \lambda^{-1}$. It is also interesting to remark that, as shown in \cite{AFR2}, the proper self-adjoint determination of \eqref{43} is somehow an intermediate case between the two we have just mentioned when we consider the behaviour of the excited state energy. There, the emergence of the second eigenvalue out of the absolutely continuous spectrum occurs exactly at  $x_0=0$.

\section{On the eigenvalues of $H_{sa}(\beta,x_0)$ as functions of $\beta<0$}

In the present Section we adopt the converse point of view with respect to that of the previous one. Now, $x_0>0$, the distance between the centres, will be held fixed and the coupling parameter $\beta<0$ will vary. Our goal is the analysis of the behaviour of both eigenvalues of $H_{sa}(\beta,x_0)$ as functions of $\beta$. 

From \eqref{37} and \eqref{38}, respectively,  we obtain the following equations that are the counterparts of \eqref{40} and \eqref{41}:
\begin{equation}\label{44}
\beta(E_0)=-\frac 2{|E_0|^{1/2}\left(1-e^{-2x_0|E_0|^{1/2}}\right)} \,,
\end{equation}
and
\begin{equation}\label{45}
\beta(E_1)=-\frac 2{|E_1|^{1/2}\left(1+e^{-2x_0|E_1|^{1/2}}\right)}\,.
\end{equation}

As a consequence of \eqref{44}, $\beta(E_0)$ is a strictly decreasing function on its domain $\left(-\infty,0\right)$ with range $\left(-\infty,0\right)$. This implies the existence of the inverse function $E_0(\beta)$, which gives the ground state energy in terms of the coupling constant for any value of $x_0>0$. Due to \eqref{45}, the same holds for $\beta(E_1)$.

\begin{figure}[htb]
\begin{center}
\includegraphics[width=0.4\textwidth]{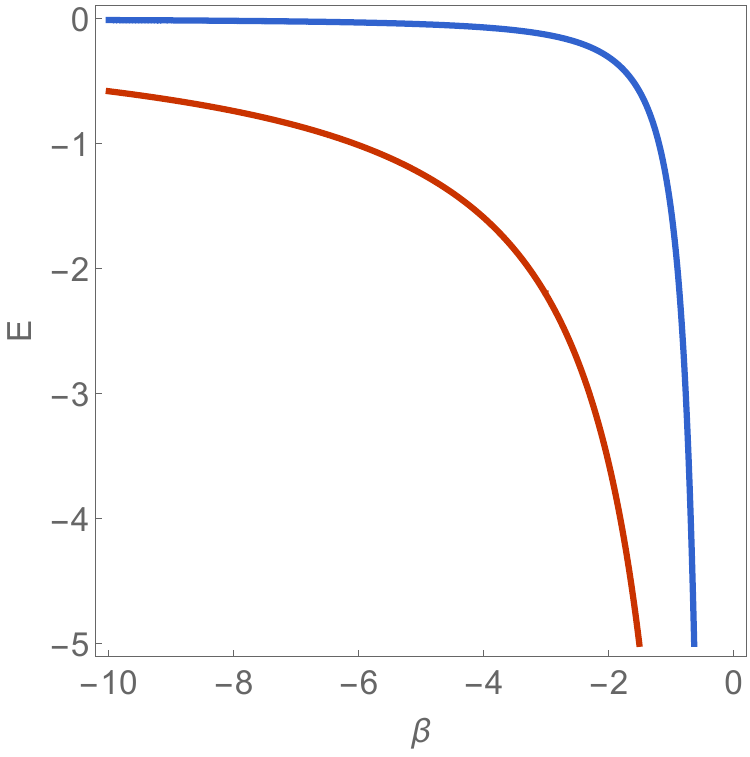}
\qquad\qquad
\includegraphics[width=0.4\textwidth]{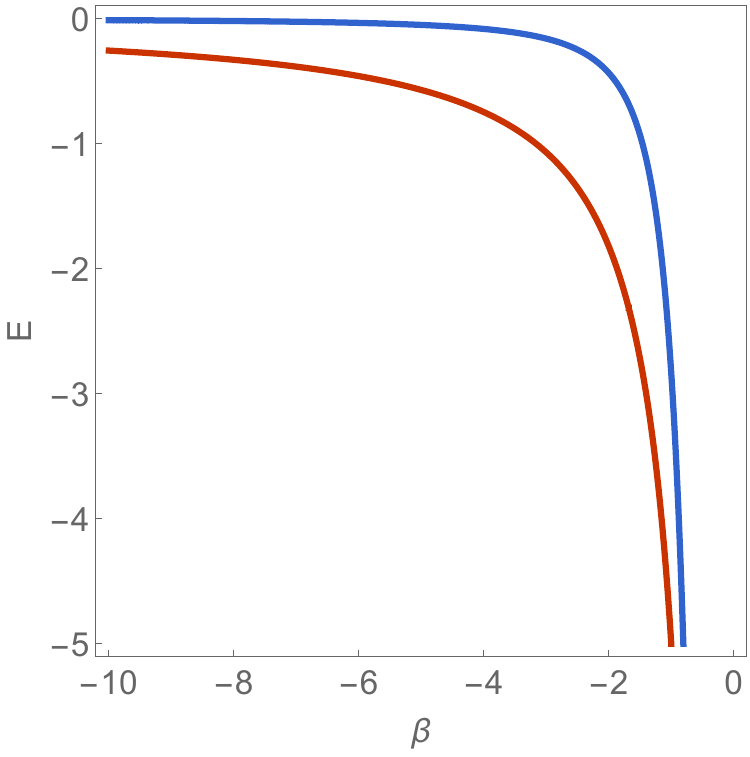}
\caption{Plot of the ground state energy $E_0(\beta)$ (red curve) and the excited state energy $E_1(\beta)$  (blue curve) as functions of $\beta$ for $x_0=0.2$ (left) and $x_0=0.5$ (right).}
\label{figure7}
\end{center}
\end{figure}

In Figure~\ref{figure7}, we see the plots of  $E_0(\beta)$ and  $E_1(\beta)$ for two different values  of $x_0$. Observe that the larger $x_0$ is, the closer the energies of both eigenstates become.

\section{Resonances}

In this Section, we show that $H_{sa}(\beta,x_0)$ has an infinite number of resonances characterised as pairs of complex poles of the resolvent of this Hamiltonian. In the momentum representation, these pairs of poles are located in the lower half of the complex plane. Each pair is symmetrically spaced with respect to the imaginary axis. 

Looking at \eqref{33}, we see that the search for complex poles of the resolvent  $R(\beta,x_0,|E|)$ is just the search for complex solutions of both \eqref{37} and \eqref{38}. Let us start with \eqref{37}. Since the energies are negative, let us replace $|E|$ by $-E$. Since we are looking for complex solutions of \eqref{37}, we write $\sqrt{- E}=-i \sqrt{E}=-i(k_1+i k_2)$, where $k_i$, $i=1,2$, are real numbers. Note that the energy $E$ always appears under a square root either in \eqref{37} or in \eqref{38}, so that this transformation is always reasonable. At the same time, we go from the energy to the momentum representation. 

Then, let us define $q_1:= 2x_0k_1$, $q_2:=2x_0k_2$ and $\alpha:= {4x_0}/{\beta}<0$. With these definitions, \eqref{37} is transformed into
\begin{equation}\label{46}
\alpha = (iq_1- q_2)\, \left(1-e^{- q_2} e^{iq_1} \right)\,.
\end{equation}
This is a complex equation, which splits into a system of two real transcendental equations, which are after some algebra:
\begin{equation}\label{47}
(q_2 + \alpha) e^{q_2} = q_2 \cos q_1 + q_1   \sin q_1  \,,
\end{equation}
and
\begin{equation}\label{48}
e^{q_2}  =  \cos q_1 - q_2  \frac{\sin q_1}{q_1}\,,
\end{equation}
where \eqref{47} and \eqref{48} correspond to the real and imaginary parts of \eqref{46}, respectively.
Observe that these equations are invariant under the transformation $q_1 \to -q_1$. This fact is important, since all possible complex solutions of \eqref{46}, and therefore of \eqref{37} appear into pairs symmetrically located with respect to the imaginary axis.

\begin{figure}[hbt]
\begin{center}
\includegraphics[width=0.325\textwidth]{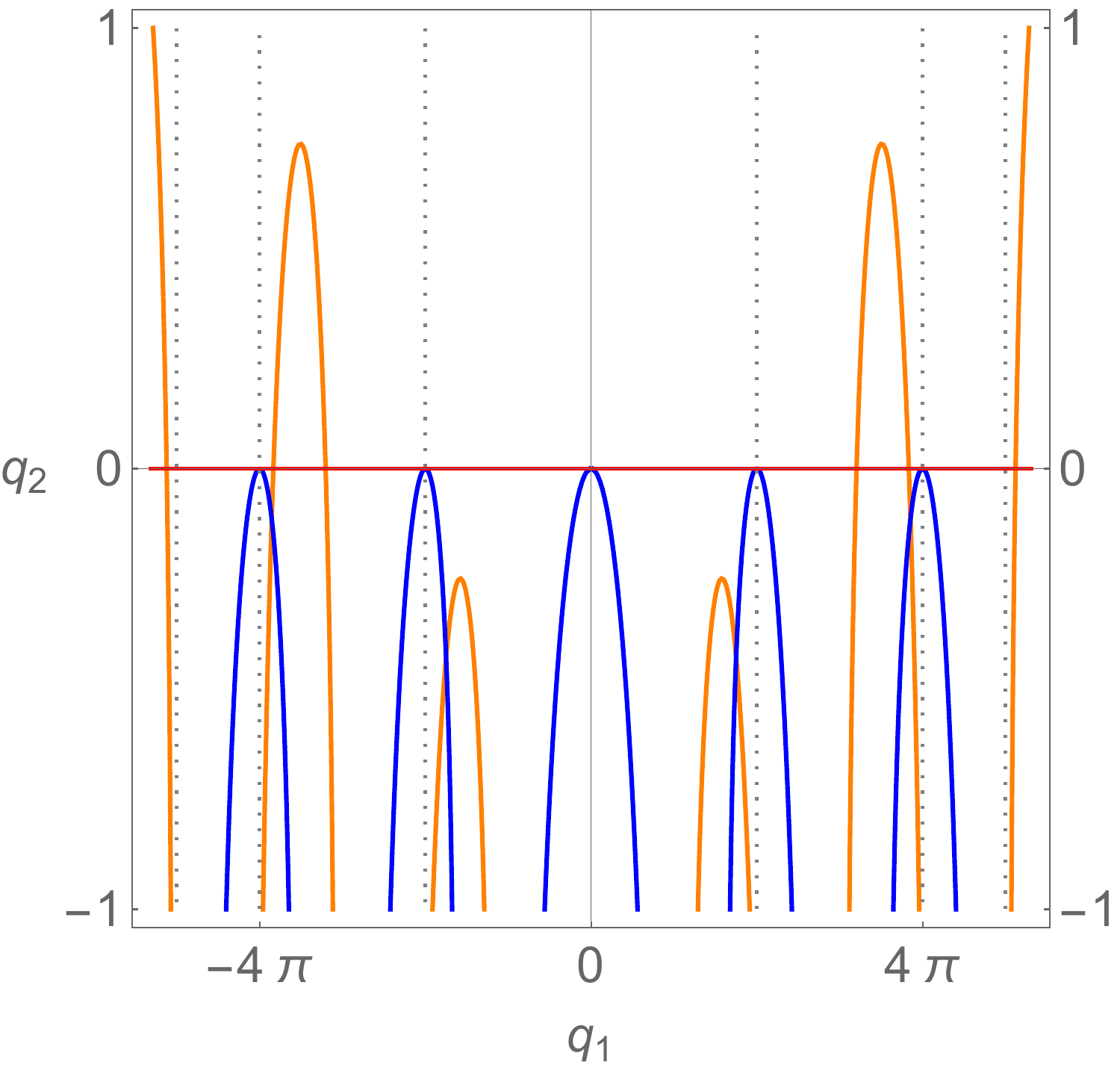}
\includegraphics[width=0.325\textwidth]{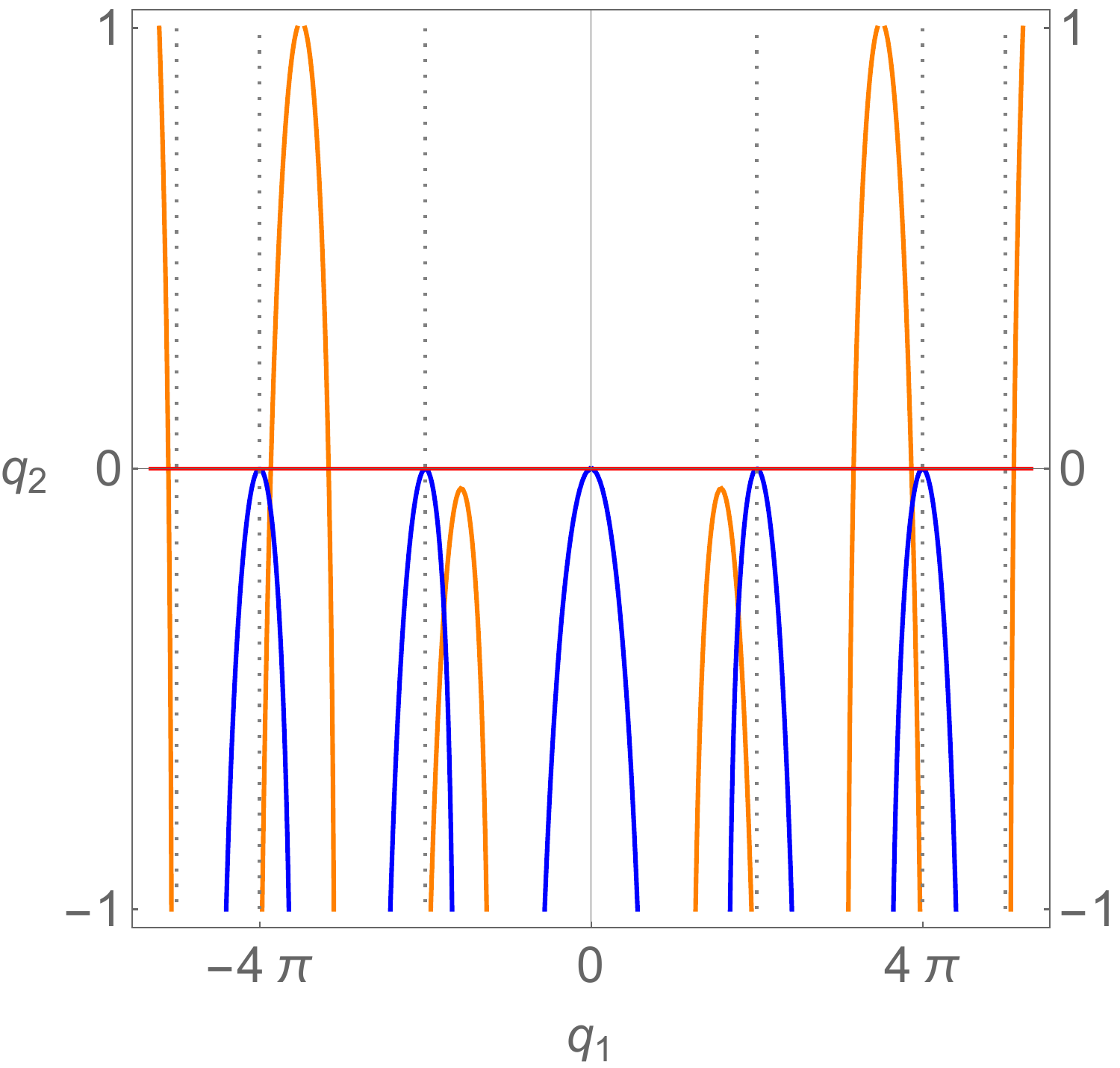}
\includegraphics[width=0.325\textwidth]{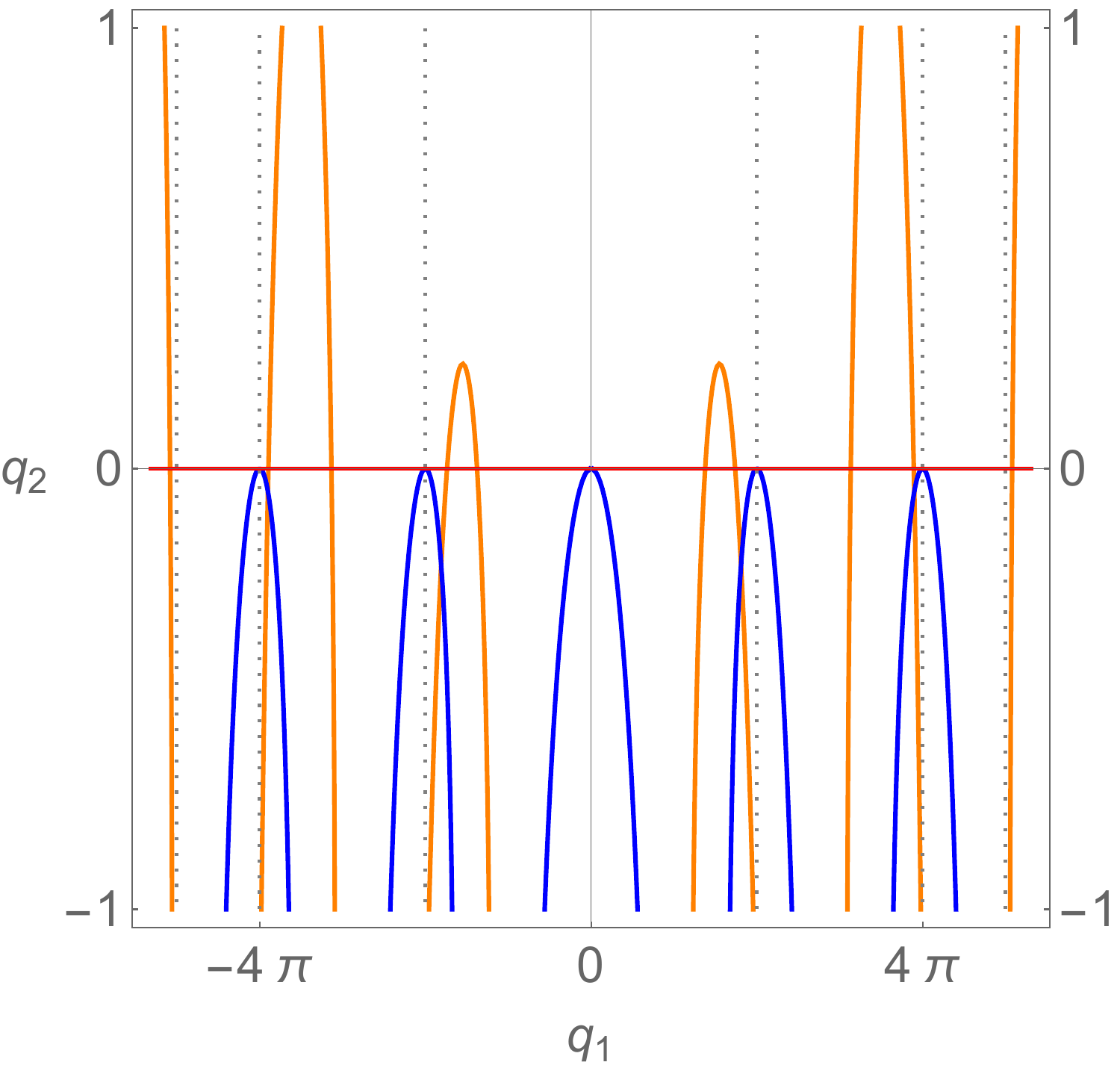}
\\ [1ex]
\includegraphics[width=0.325\textwidth]{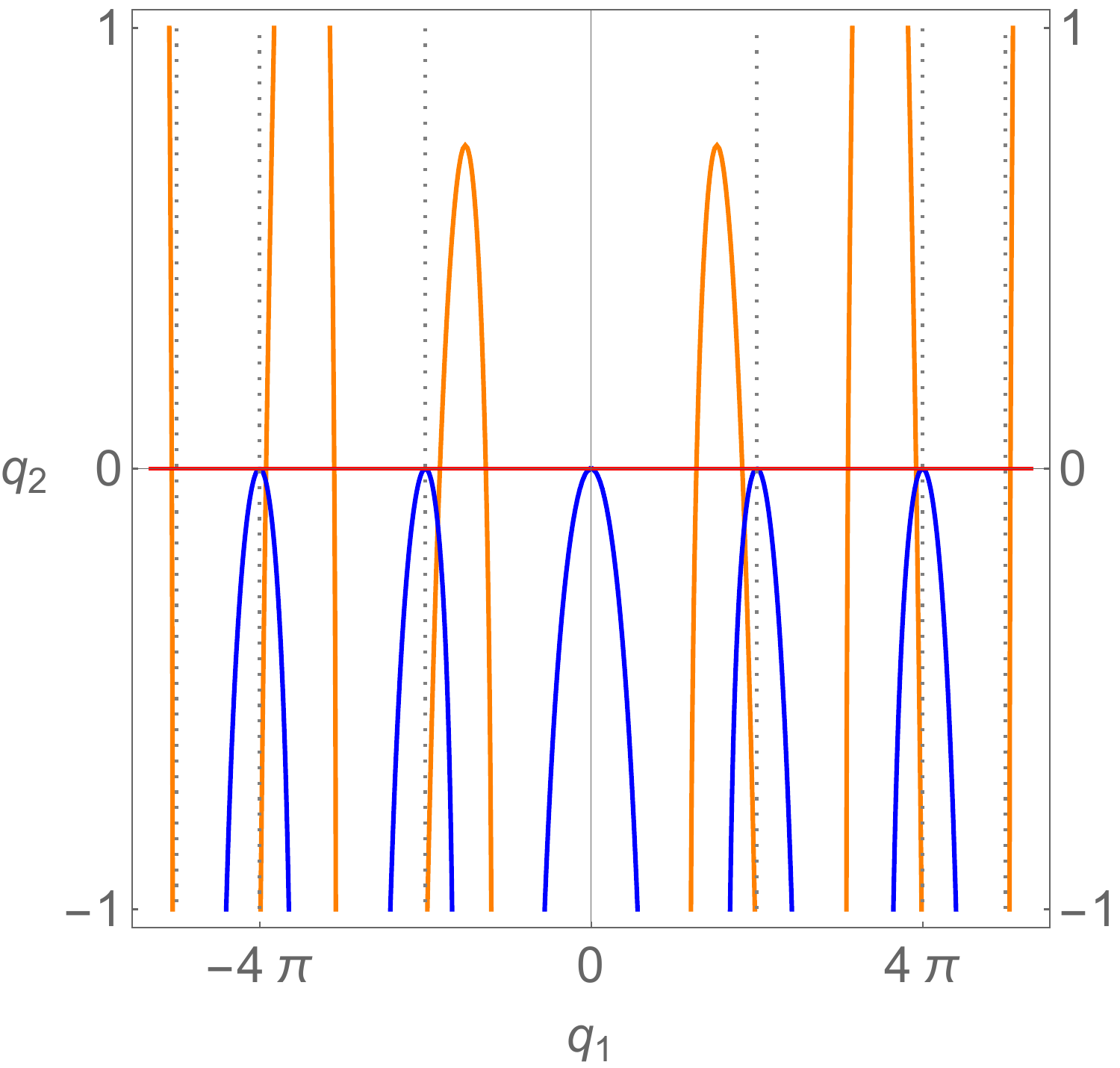}
\includegraphics[width=0.325\textwidth]{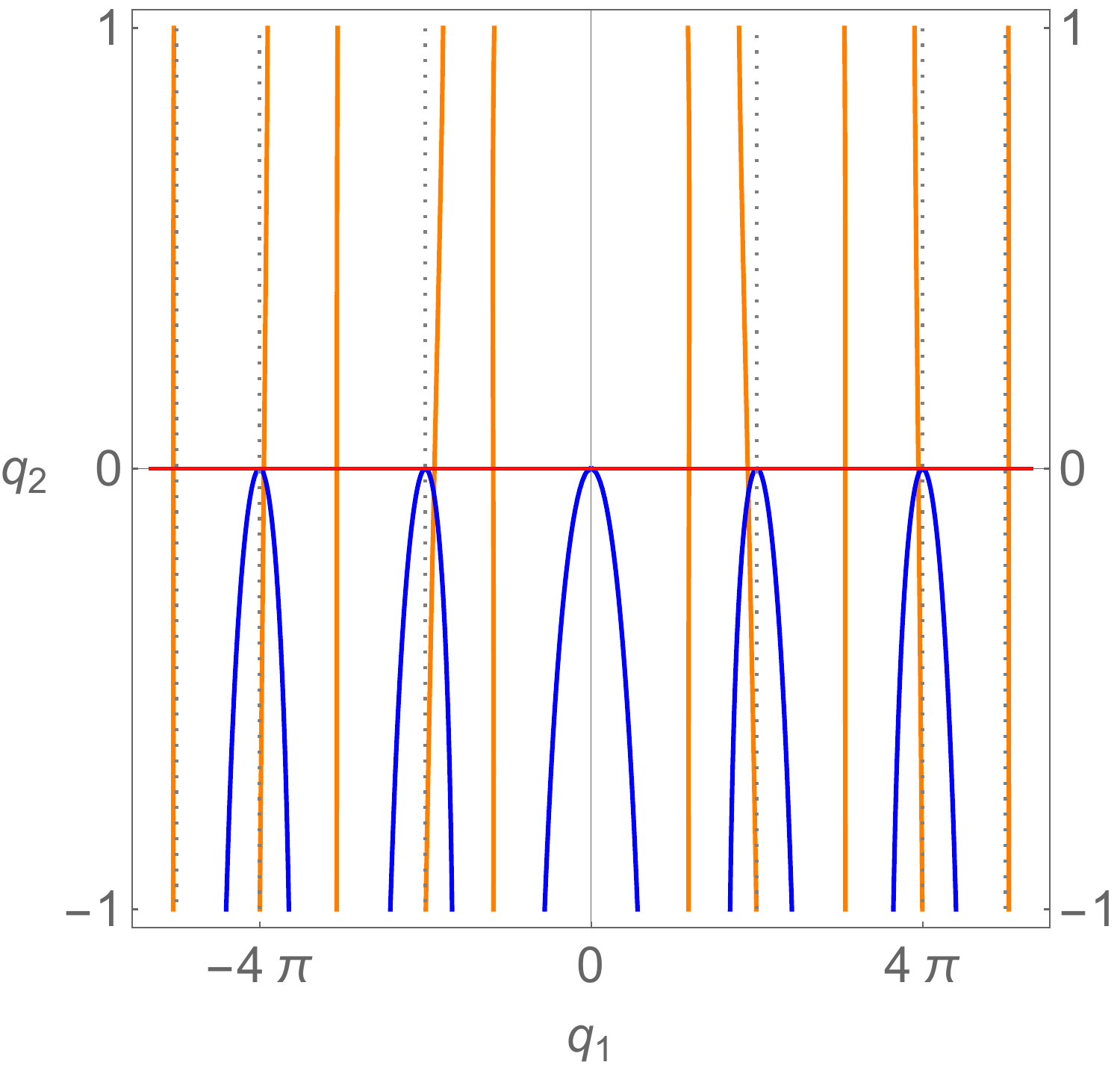}
\includegraphics[width=0.325\textwidth]{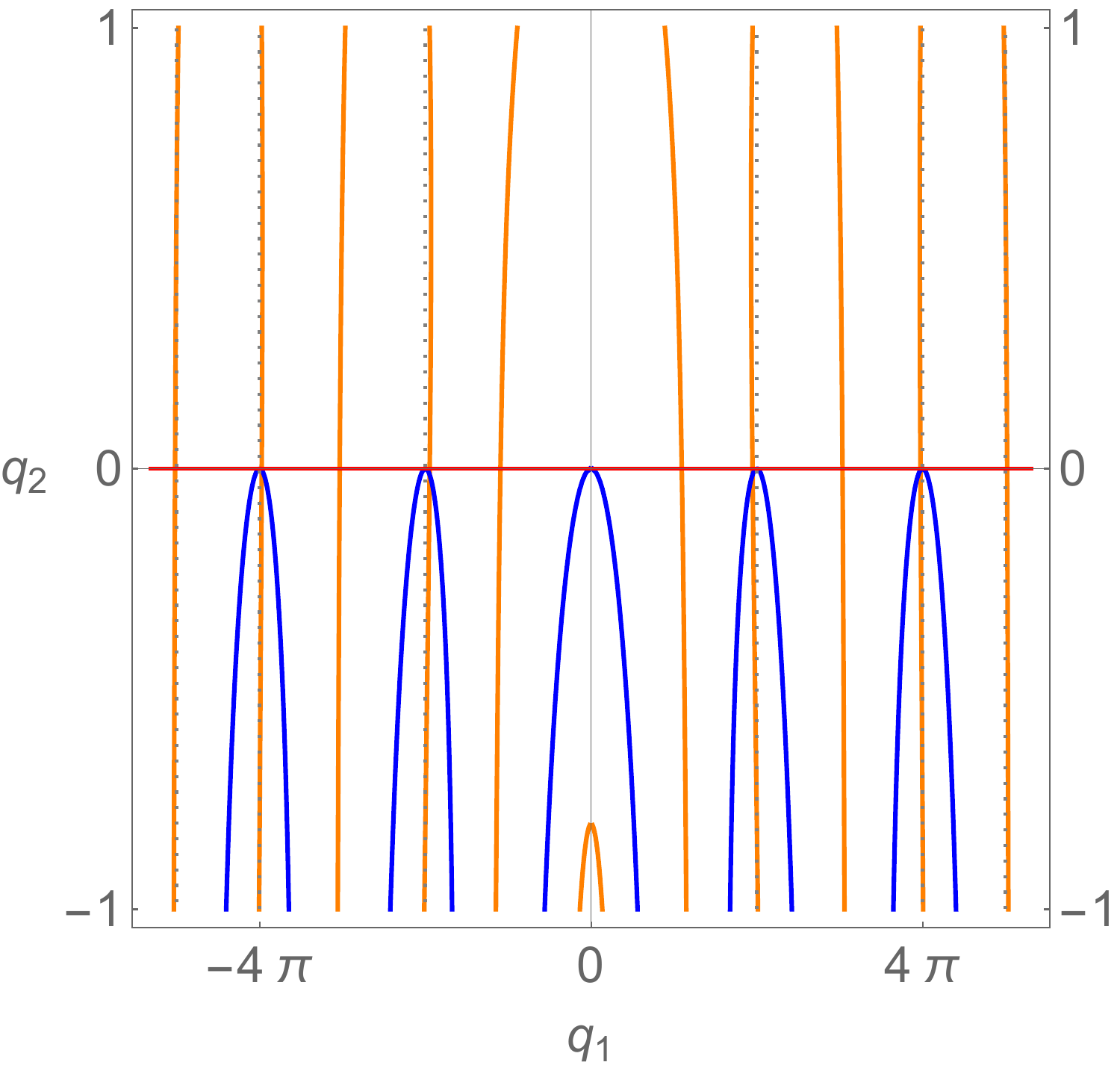}
\caption{Plot of the resonances associated to equations \eqref{47}--\eqref{48}, which coincide with the intersection of the  orange and blue curves.  From left to right, and from up to down,  $\alpha= -6,-5,-4,-3,-2,-1$.}
\label{figuregroundresonances}
\end{center}
\end{figure}

Then, the intersections of curves \eqref{47} and \eqref{48}, orange and  blue, respectively in Figure~\ref{figuregroundresonances}, give the resonance poles on the momentum plane. Observe that these poles come into pairs symmetrically spaced with respect to the imaginary axis and have a negative imaginary part. According to a general theory \cite{NUS,KKH}, these pairs of poles correspond to scattering resonances. In Figure~\ref{figuregroundresonances}, we show the location for the first two resonances for various values of $\alpha$. On the graphics one sees that the larger the value of $q_1$ for a resonance is, the closer the resonance pole to the $q_2=0$ axis will be.

Once we have obtained the resonances as the complex zeroes of \eqref{37}, we may repeat the steps with \eqref{38}. Here, the counterpart of \eqref{47} is
\begin{equation}\label{49}
(q_2+\alpha) e^{q_2} = - \left( q_2  \cos q_1 + q_1  \sin q_1  \right)\,,
\end{equation}
and of \eqref{48}:
\begin{equation}\label{50}
e^{q_2}  = - \left( \cos q_1 - q_2   \frac{\sin q_1}{q_1}\right)\,.
\end{equation}

\begin{figure}[htb]
\begin{center}
\includegraphics[width=0.325\textwidth]{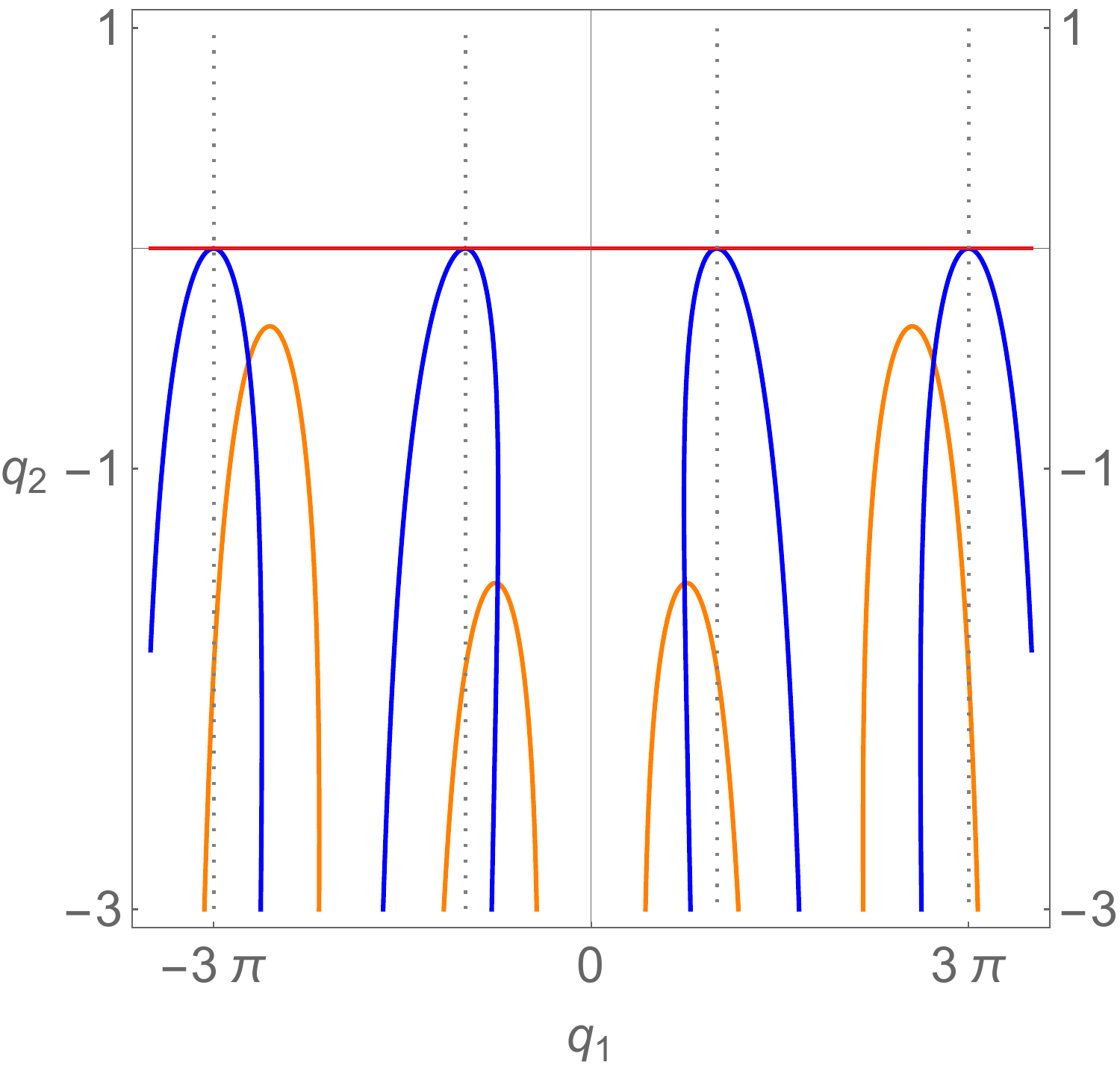}
\includegraphics[width=0.325\textwidth]{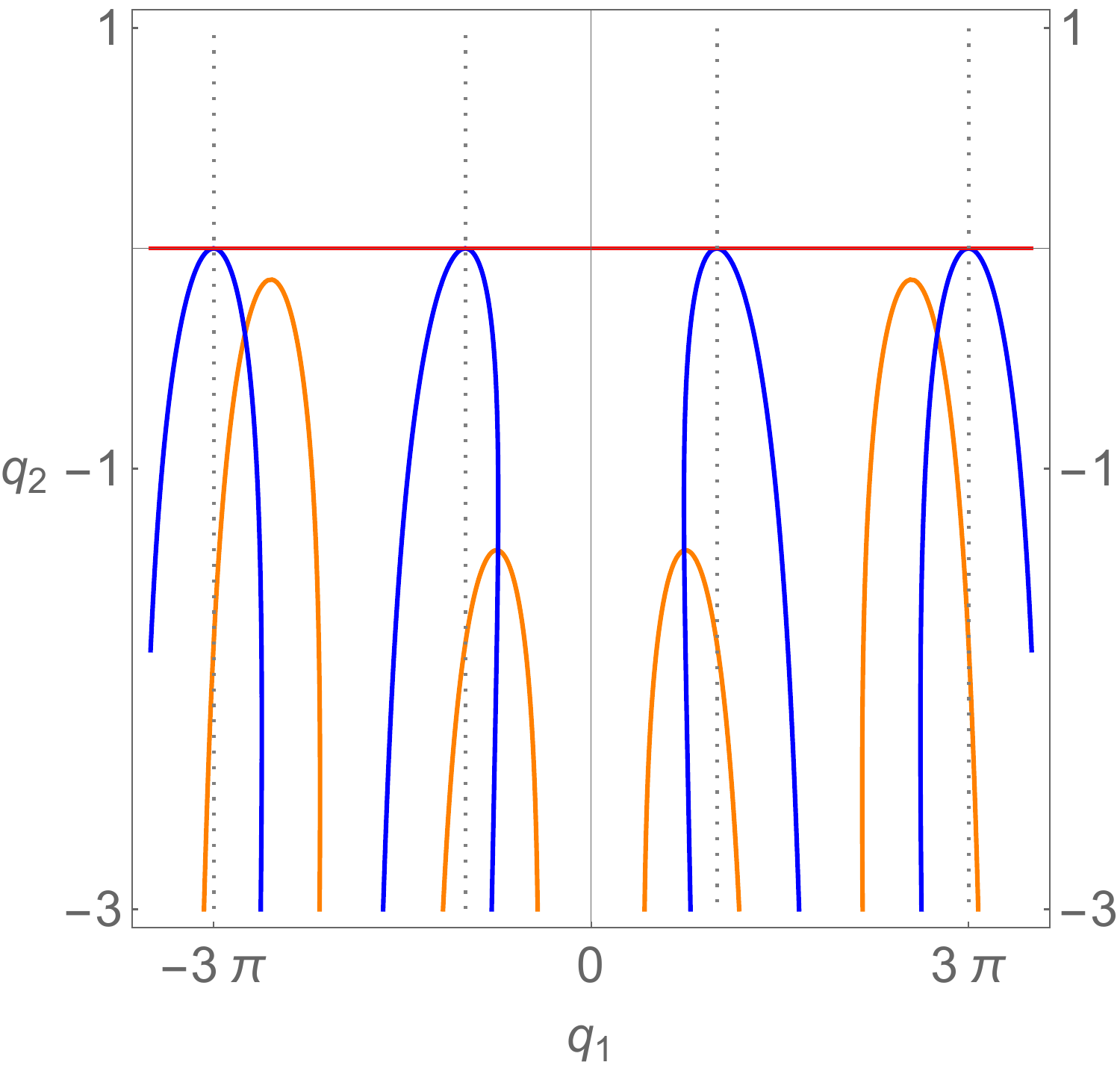}
\includegraphics[width=0.325\textwidth]{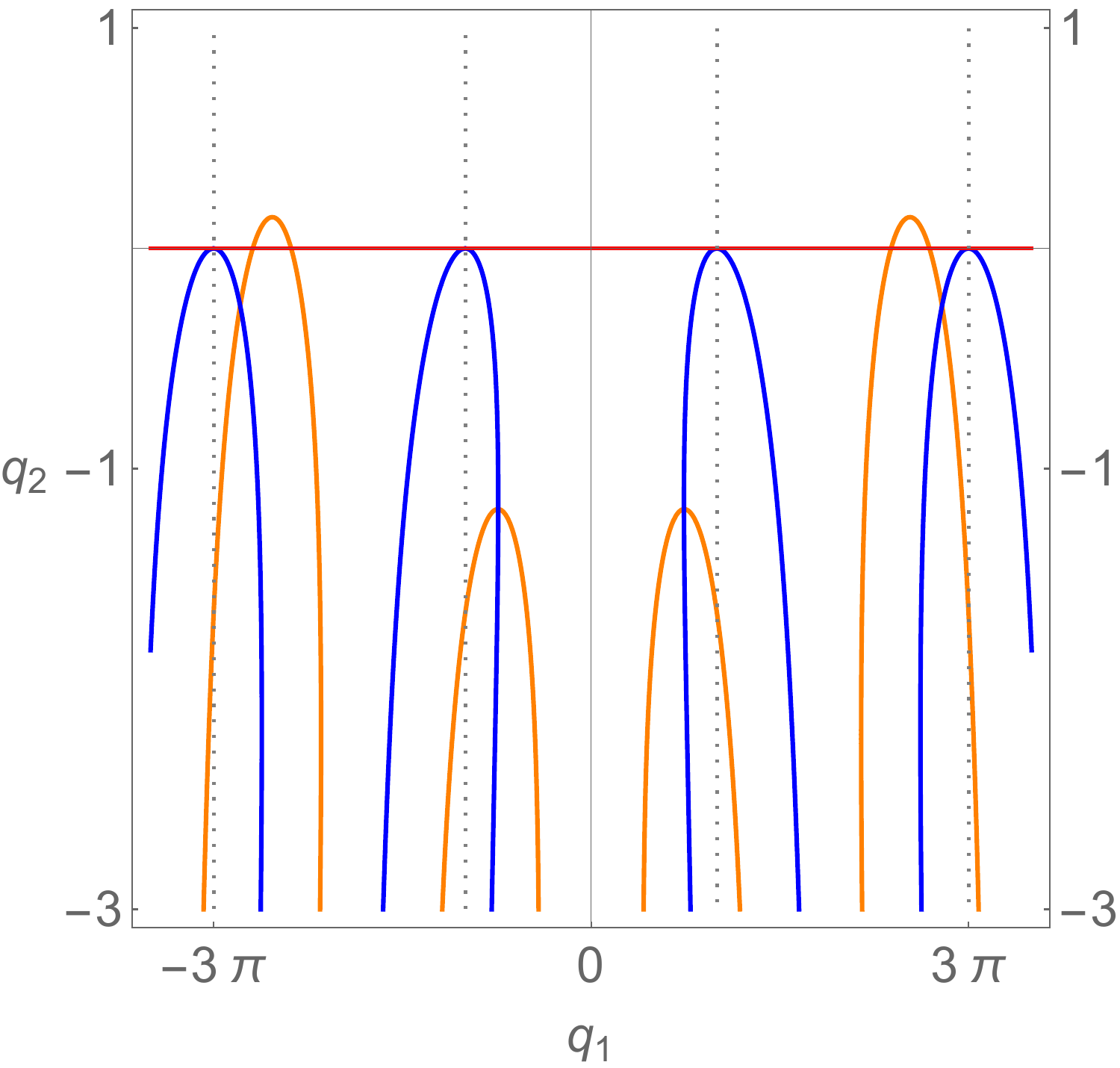}
\\ [1ex]
\includegraphics[width=0.325\textwidth]{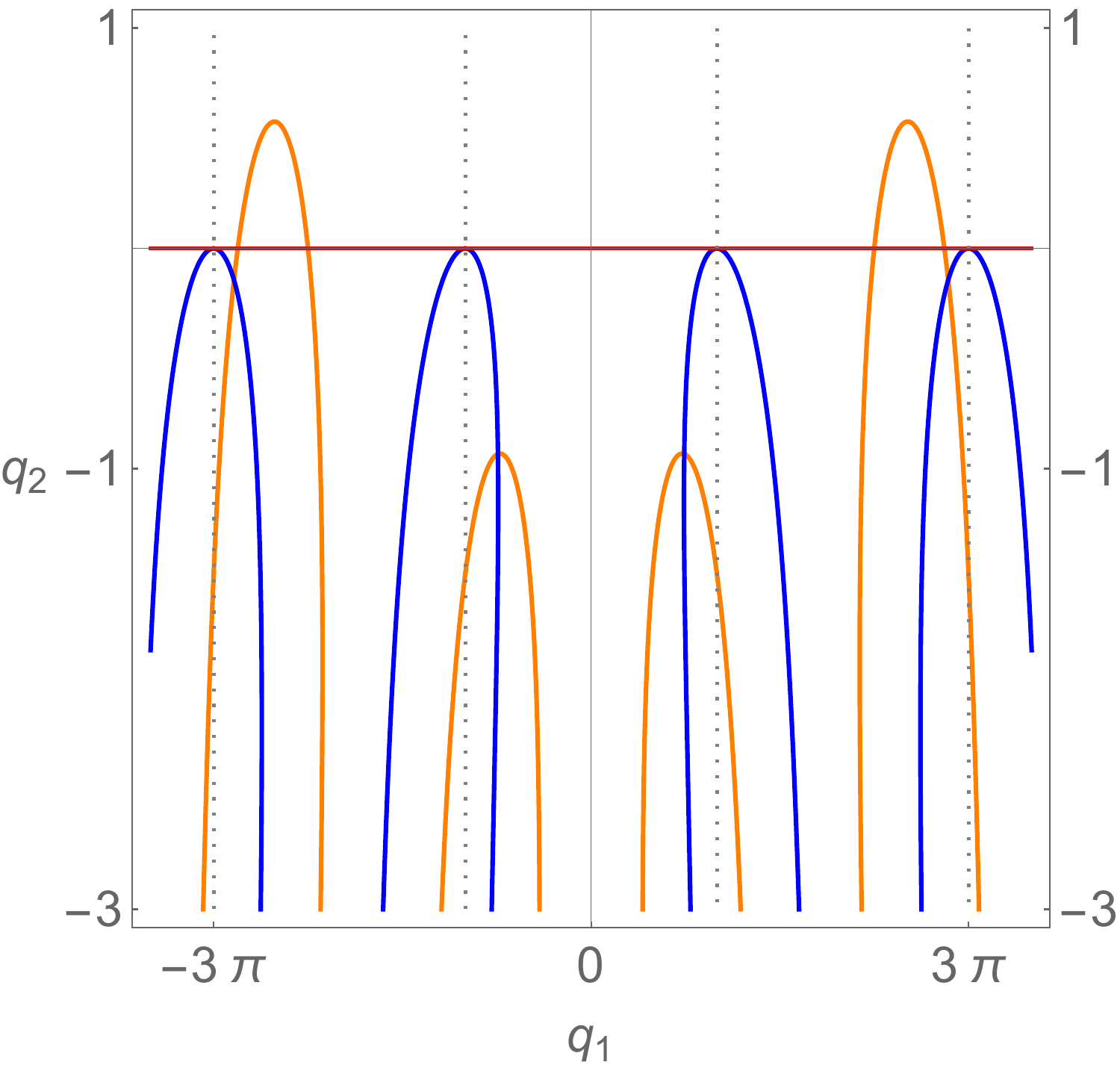}
\includegraphics[width=0.325\textwidth]{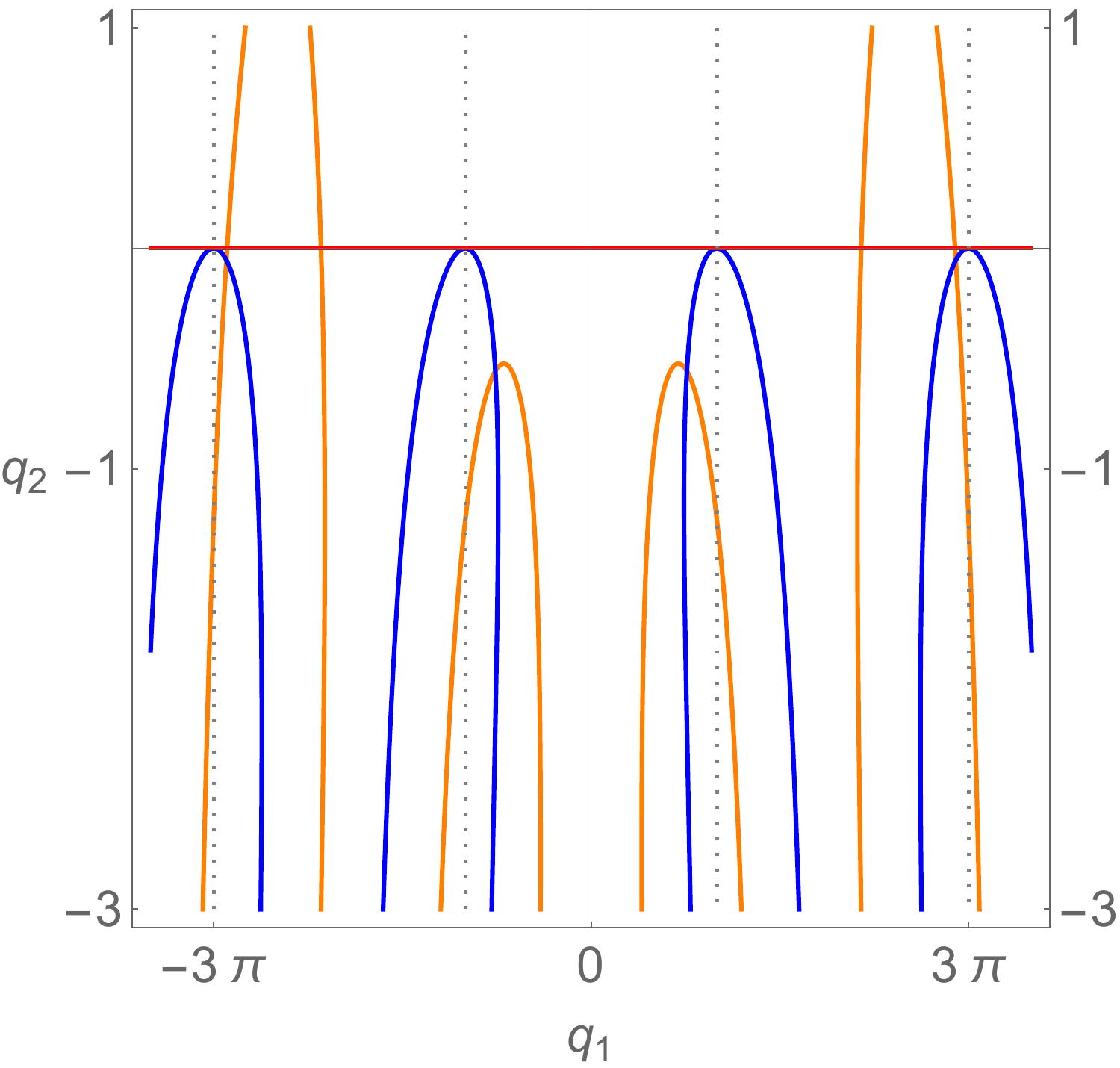}
\includegraphics[width=0.325\textwidth]{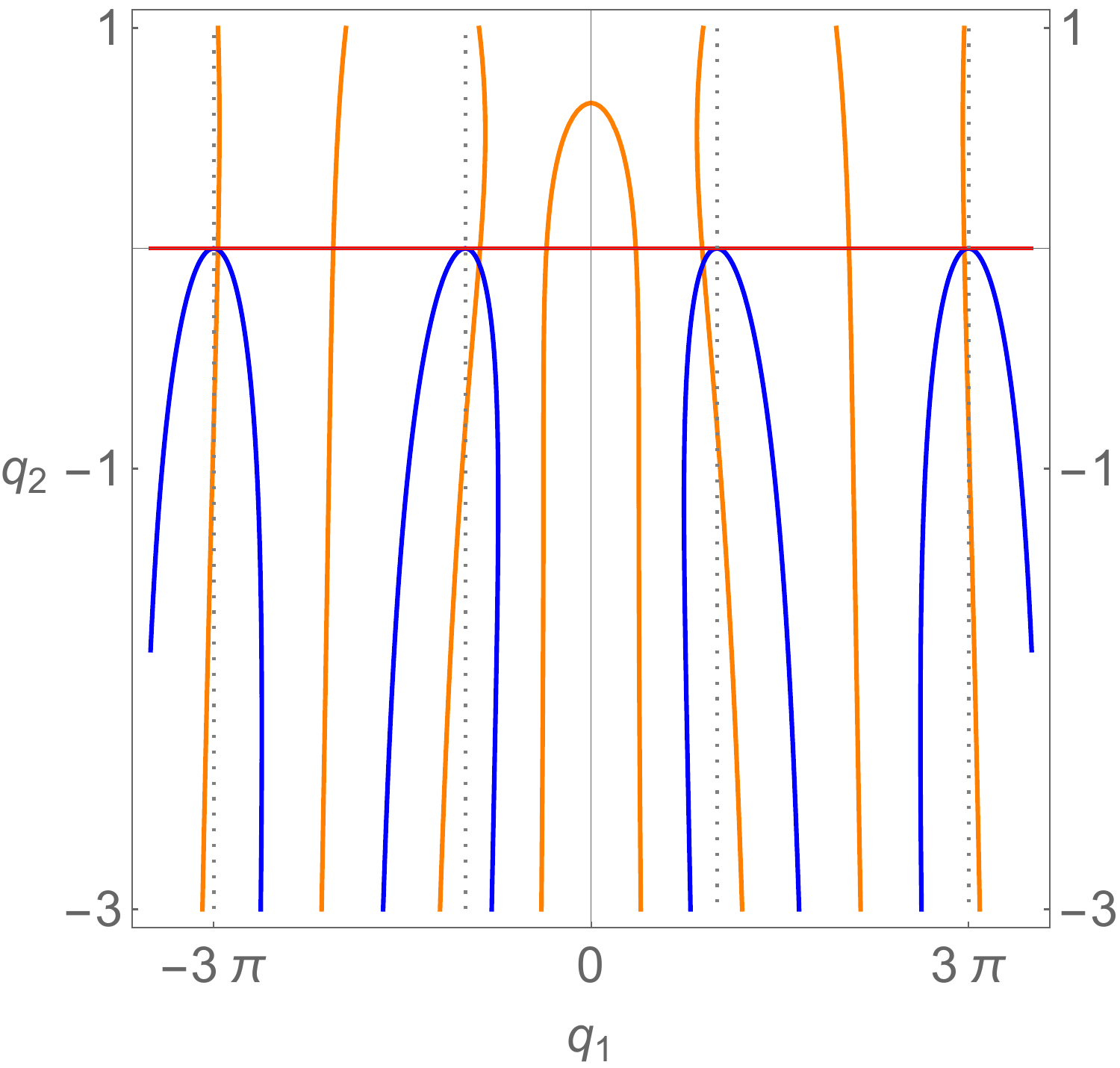}
\caption{Plot of the resonances associated to equations \eqref{49}--\eqref{50}, which coincide with the intersection of the  orange and blue curves.  From left to right, and from up to down,  $\alpha= -11,-9,-7,-5,-3,-1$.}
\label{figureexcitedresonances}
\end{center}
\end{figure}

In Figure~\ref{figureexcitedresonances}, we depict the  first two resonance poles for this second pair of equations as the intersections of the blue and orange curves, for various values of $\alpha$. Curves \eqref{50} and in \eqref{49} are depicted in blue and orange, respectively. Observe that in boths cases the behaviour of such resonance poles is quite similar. 

At this stage we want to show that all resonance poles on the momentum plane $(q_1,q_2)$ lie in the open lower half plane. In order to achieve this, we need only prove that in the curves \eqref{48} and \eqref{50} one necessarily has that $q_2\le 0$. In fact from both \eqref{48} and \eqref{50}, it follows that
\begin{equation}\label{51}
0< e^{q_2} \le |\cos q_1| + |q_2 | \, \left| \frac{ \sin q_1}{q_1}  \right| \le 1 + |q_2| < e^{|q_2|}\,.
\end{equation}
The latter inequality is strict for $q_2 \ne 0$, so that if $q_2$ were positive, we should have $e^{q_2} < e^{q_2}$, which is a nonsense. Therefore, $q_2\le 0$ for all curves \eqref{48} and \eqref{50}, so that they lie in the lower half plane of the plane $(q_1,q_2)$. 

In addition, if $q_2=0$ in \eqref{48} (real axis), then, $q_1= 2\pi n$, $n=0,\pm 1,\pm 2, \dots$. If $q_2=0$ in \eqref{50}, then, $q_1=(2n-1)\pi$, $n=0,\pm 1,\pm 2, \dots$. It is not difficult to show that all these points are relative maxima of \eqref{48} and \eqref{50} respectively. These facts are clearly shown in Figures~\ref{figuregroundresonances} and \ref{figureexcitedresonances}. 

Finally, we note that this model does not show either anti-bound states, also called virtual states \cite{NUS,ID,GKN}, or redundant states \cite{MA,MM,GHKN}.

\section{On the behaviour of $H_{sa}(\beta,x_0)$ as $x_0\to 0^+$}

Throughout this section $\beta \ne 0$ will be assumed to be fixed. In order to obtain the limit of $H_{sa}(\beta,x_0)$ as the distance between both centres vanishes, we are going to study this limit in the resolvent equation \eqref{33}, where we consider each term separately. First of all, note that as $x_0\to 0^+$, we have that
\begin{equation}\label{52}
\frac 1{\pi\, \left[\frac 1{\beta} + \frac {|E|^{1/2}}{2}\, \left(1-e^{-2x_0|E|^{1/2}} \right) \right]} \to \frac {\beta}{\pi}\,.
\end{equation}

Then, observe that for any $x_0>0$, we have the following upper bound:
\begin{equation}\label{53}
\left \|\frac {p\,\sin x_0 p}{p^2+|E|}\right \| _2^2 = \int_{-\infty}^{\infty}\, \frac {p^2\,\sin^2x_0 p}{\left(p^2+|E|\right)^2}\, dp \le \int_{-\infty}^{\infty}\, \frac {p^2}{\left(p^2+|E|\right)^2}\, dp=\frac {\pi}{2|E|^{1/2}}\,,
\end{equation}
where $||-||_2$ is the norm on $L^2(\mathbb R)$. Then, if we apply the Lebesgue dominated theorem to the first integral on \eqref{53}, we can conclude that
\begin{equation}\label{54}
\lim_{x_0\to 0^+} \left \|\frac {p\,\sin x_0 p}{p^2+|E|}\right \| _2 =0\,.
\end{equation}

By looking at the second term in \eqref{33}, we note that it is a rank one operator acting on the subspace of even functions. Due to \eqref{54}, it is not difficult to show that its trace norm vanishes as $x_0\to 0^+$. The proof is the following: Let $||-||_{T_1}$ be the trace norm. Then,
\begin{eqnarray}\label{55}
\left \| \left| \frac {p\, \sin x_0 p}{p^2+|E|} \right> \left< \frac {p\, \sin x_0 p}{p^2+|E|} \right| \right \| _{T_1} = \left \|\frac {p\,\sin x_0 p}{p^2+|E|}\right \| _2^2 \to 0\,,
\end{eqnarray}
as $x_0\to 0^+$.

Then, we turn our attention to the last term in the resolvent in \eqref{33}, which is again a rank one operator. First of all, as $x_0\to 0^+$, we have the following limit
\begin{equation}\label{56}
\frac 1{\pi\, \left[\frac 1{\beta} + \frac {|E|^{1/2}}{2}\, \left(1+e^{-2x_0|E|^{1/2}} \right) \right]} \to \frac 1{\pi\, \left(\frac 1{\beta} + |E|^{1/2} \right)}\,.
\end{equation}

In addition, for arbitrary $f,g \in L^2(\mathbb R)$, we have the following limit as $x_0\to 0^+$:
\begin{equation}\label{57}
\left(f, \left| \frac {p\, \cos x_0 p}{p^2+|E|} \right> \left< \frac {p\, \cos x_0 p}{p^2+|E|} \right|\, g\right) \to \left(f, \left| \frac {p}{p^2+|E|} \right> \left< \frac {p}{p^2+|E|} \right|\, g\right)\,.
\end{equation}

The results given by equations \eqref{56} and \eqref{57}, show the convergence, in the weak topology of bounded operators, of the third term of \eqref{33} to
\begin{equation}\label{58}
\frac 1{\pi\, \left(\frac 1{\beta} + |E|^{1/2} \right)} \left| \frac {p}{p^2+|E|} \right> \left< \frac {p}{p^2+|E|} \right|\,.
\end{equation}

Furthermore, this convergence actually holds in the trace norm topology. Using again the dominated convergence theorem, taking the limit as $x_0\to 0^+$, we have the following:
\begin{eqnarray}\label{59}
&&
\left \| \left| \frac {p\, \cos x_0 p}{p^2+|E|} \right> \left< \frac {p\, \cos x_0 p}{p^2+|E|} \right| \right \| _{T_1} = \left \|\frac {p\,\cos x_0 p}{p^2+|E|}\right \| _2^2 = \int_{-\infty}^{\infty}\, \frac {p^2\,\cos^2x_0 p}{\left(p^2+|E|\right)^2}\, dp \nonumber \\ [1.2ex]
&& \hskip2.3cm
\to \int_{-\infty}^{\infty}\, \frac {p^2}{\left(p^2+|E|\right)^2}\, dp = \left \|\frac {p}{p^2+|E|}\right \| _2^2 = \left \| \left| \frac {p}{p^2+|E|} \right> \left< \frac {p}{p^2+|E|} \right| \right \| _{T_1}\,,\qquad\,
\end{eqnarray}
which implies the above-mentioned convergence in the trace norm topology as a consequence of Theorem 2.21 in \cite{TraceId}.

Therefore, as $x_0\to 0^+$, we have:
\begin{eqnarray}\label{60}
\left \| \left[H_{sa}(\beta,x_0)+ |E|\, \right]^{-1}-\left(p^2+|E|\right)^{-1}-\frac 1{\pi\, \left(\frac 1{\beta} + |E|^{1/2} \right)} \left| \frac {p}{p^2+|E|} \right> \left< \frac {p}{p^2+|E|} \right| \right \|_{T_1} \to 0\,.
\end{eqnarray}

At this stage, in principle, we should prove that the limiting operator
\begin{equation}\label{61}
\left(p^2+|E|\right)^{-1}+\frac 1{\pi\, \left(\frac 1{\beta} + |E|^{1/2} \right)} \left| \frac {p}{p^2+|E|} \right> \left< \frac {p}{p^2+|E|} \right|
\end{equation}
is the resolvent of a self-adjoint operator. However, by comparing the second term in \eqref{61} with (5.8) in \cite{FR} (taking account of the fact that in \cite{FR} the negative sign in front of $\beta$ had been introduced by default), it is almost immediate to realise that, for any $\beta<0$ and $E<0$,
\begin{equation}\label{62}
\left(p^2+|E|\right)^{-1}+\frac 1{\pi\, \left(\frac 1{\beta} + |E|^{1/2} \right)} \left| \frac {p}{p^2+|E|} \right> \left< \frac {p}{p^2+|E|} \right|\, =\left(\Xi_{2\beta}+|E|\, \right)^{-1}\,,
\end{equation}
where, following \cite{AGHH}, $\Xi_{2\beta}$ represents $-d^2/dx^2$ on $W^2_2(\mathbb R/\{0\})$ with the following two-sided boundary conditions at the origin: $\psi'(0^+) = \psi'(0^-)$ and $\psi(0^+) -\psi(0^-) = 2\beta \psi'(0)$, for any $\psi(x)$ in the domain of $\Xi_{2\beta}$. These are exactly the conditions that determine the nonlocal interaction $2\beta\,\delta'(x)$, so that $\Xi_{2\beta}$ is the self-adjoint determination of the heuristic expression $-\frac {d^2}{dx^2}-2\, \lambda \, |\delta'\rangle\langle\delta'|$. 

In conclusion, the self-adjoint Hamiltonian $H_{sa}(\beta,x_0)$ converges in the norm resolvent sense to the self-adjoint operator $\Xi_{2\beta}$ as $x_0 \to 0^+$. We may say that, as the distance between the centres vanishes, these two identically attractive $\delta'$-interactions smoothly coalesce and become a single attractive $\delta'$-interaction supported at the origin with strength $2\beta$. 

With regard to the spectrum of $\Xi_{2\beta}$, we can say that, for any $\beta<0$, this operator has one simple negative eigenvalue, $E_0(\beta)=-1/\beta^2$ and its absolutely continuous spectrum is $[0,\infty)$. 

All these results were expected after looking at the behaviour of the spectral curves as functions of $x_0$ depicted in Figures~\ref{figure1}--\ref{figure3}. In fact, while the lower curve, corresponding to the ground state as a function of $x_0$ diverges negatively in the limit $x_0 \to 0^+$, the excited state always approaches the value $-1/\beta^2$. We may say that since {\it the principle of noncontraction of the spectrum} holds under {\it norm} resolvent convergence \cite{RSI}, the value $-1/\beta^2$ {\it may not abruptly} disappear from the spectrum of the limiting operator. It is worth stressing that this principle does not hold under {\it strong} resolvent convergence, which only ensures that the spectrum of the limiting operator may not suddenly expand \cite{RSI}. 

We may summarise the latest results as follows:

\medskip

\noindent
{\bf Theorem 2.} {\it  For any fixed value of $\beta \neq 0$, the self-adjoint Hamiltonian $H_{sa}(\beta,x_0)$ whose resolvent is given by $R(\beta,x_0,|E|)$ in \eqref{33}, for any $E<0, x_0>0$, converges in the norm resolvent sense to the self-adjoint Hamiltonian $\Xi_{2\beta}$, namely the negative Laplacian with the well-known $\delta'$-conditions   \eqref{10} with coupling constant $2\beta$ at the origin, as $x_0 \to 0^+$.}

\section{Some further discussions and concluding remarks}

In previous articles where the free Hamiltonian has been either $H_0=-d^2/dx^2$ (see \cite{FR}) or $H_0= \frac12 \left[- \frac{d^2}{dx^2}   + x^2 \right]$ (see \cite{FR1}), we have studied the perturbation given by $-\lambda \left[\delta(x+x_0)+\delta(x-x_0)\right]$, $\lambda >0$.  As the half-distance between the two centres $x_0 \to 0^+$, each Hamiltonian converges, in the norm resolvent sense, to the respective Hamiltonian $H_0-2\lambda \,\delta(x)$. Note that in the limit the coupling constant gets doubled. This is somehow an expected result, as the one-dimensional $\delta$-perturbation is not {\it too singular} since it is an infinitesimally small perturbation of either free Hamiltonian, as a consequence of the KLMN Theorem \cite{RSII}. This implies that the coupling constant renormalisation is not required in this case and the one-dimensional $\delta$ behaves {\it essentially} like a short range smooth potential.

A completely different situation arises with the $\delta$-perturbation of the free Salpeter Hamiltonian \cite{AK1,AFR2,HSW,EGU} given by
\begin{equation}\label{63}
H_0 = \sqrt{-\frac{d^2}{dx^2} + m^2}
\end{equation}

In this case the KLMN theorem does not hold, so that the one-dimensional Dirac distribution is no longer infinitesimally small with respect to $H_0$. Therefore, the renormalisation of the coupling constant is needed in order to define rigorously the self-adjoint operator, $H(\beta,x_0)$, making sense of the heuristic expression $H_0\,-\lambda \left[\delta(x+x_0)+\delta(x-x_0)\right]$, with $H_0$ as in \eqref{63}, see also \eqref{43}. Here $\beta$ is the coupling parameter arising from the renormalisation procedure.  It has been rigorously proved \cite{AFR2} that in the limit as $x_0 \to 0^+$, the self-adjoint operator $H(\beta,x_0)$ does not converge to $H(2\beta,0)$. Thus, the two point interactions do not merge smoothly at the origin. This pathology has a cure, that is to say the renormalised strength parameter is to be made dependent on the distance between the centres, $\beta \equiv \beta(x_0)$. Then, one shows that $H(\beta(x_0),x_0)$ converges in the norm resolvent sense to $H(2\beta,0)$, thus making the smooth merging of the two point interactions possible.

A similar situation occurs for singular perturbations either of $H_0=-\Delta$ or $H_0=\frac 12 [-\Delta +|\mathbf x|^2]$ in two dimensions, with centres at $(-x_0,0)$ and $(x_0,0)$, or three dimensions with centres at $(-x_0,0,0)$ and $(x_0,0,0)$ \cite{AFR3,FPR,AFR,AFR1}. 

In view of the above remarks, it is slightly bewildering that two extremely singular $\delta'$-interactions, which do require a renormalisation  procedure in order to be rigorously defined as perturbations of the free Hamiltonian $H_0=-\frac {d^2}{dx^2}$, can coalesce smoothly as the distance between the centres vanishes, as shown in the present manuscript. We propose a possible explanation for this difference: as a matter of fact, what really matters is the behaviour of $E_1(x_0)$ in a right neighbourhood of  $x_0=0$, due to the principle of noncontraction of the spectrum under norm resolvent convergence.  The fact that $H(2\,\beta,0)$ and $H_{sa}(\beta,x_0)$ are defined by renormalisation does not really matter, because the symmetric ground state disappears in the limit, differently from the Salpeter Hamiltonian or the 2D/3D Hamiltonians studied in the aforementioned articles.

The latter fact leads us to point out a rather remarkable phenomenon exhibited by this simple one-dimensional model: while for any $x_0>0$, the ground state wave function is clearly symmetric, at the critical value $x_0=0$ the wave function of the unique bound state becomes antisymmetric. It may be worth noting that this spectral phenomenon is somewhat reminiscent of the one described by Klaus in \cite{K3} dealing with the Hamiltonian with an attractive Coulomb potential in one dimension and its approximants involving a cutoff. It is worth stressing that Klaus' rigorous functional analytic approach represented a major contribution toward a better understanding of this model.

Remarkably, this kind of symmetry reversal also occurs when:

i.)  The coupling parameter of an attractive nonlocal $\delta'$-interaction centred at the origin, perturbing the Hamiltonian of the one-dimensional harmonic oscillator, exceeds the critical value $\beta_0=\frac {\Gamma(1/4)}{2\Gamma(3/4)}\approx 1.47934$, as shown in \cite{FGGNR}. See also \cite{AFR4,AFR5}. 

ii.) The coupling parameter of an attractive nonlocal $\delta'$-interaction centred at the origin, perturbing the Hamiltonian of the one-dimensional conic or V-shaped oscillator, exceeds the critical value $\beta_0=-\frac {Ai(0)}{Ai'(0)}\approx  1.37172$, as shown in \cite{FGGNR} (see \cite{FGGN} as well). 

Both models exhibit the phenomenon called {\it level crossing of eigenvalues},  thoroughly discussed in  \cite{FGGNR}, which induces the double degeneracy of the ground states for critical values of the coupling constant, as given before. As shown in detail in\cite{AGHH}, this double degeneracy also manifests itself when the operator $\left(-\frac {d^2}{dx^2}\right)_{\theta}$,  the one-dimensional negative Laplacian with the well-known $\theta$-boundary conditions acting on on $L^2[-a/2,a/2]$, is perturbed by an attractive nonlocal $\delta'$ interaction supported at the origin. See \cite{RSIV} for a definition of the operator $\left(-\frac {d^2}{dx^2}\right)_{\theta}$ and its role as a fibre of $-\frac {d^2}{dx^2}$, and \cite{AGHH} for a self-adjoint determination of the heuristic Hamiltonian  $\left(-\frac {d^2}{dx^2}\right)_{\theta}\,+\, \lambda \, \delta'(x)$ and its role as a fibre of the negative Laplacian in one dimension decorated with a periodic array of nonlocal $\delta'$ interactions. 

As to our concluding remarks, we see how apparently simple models provide both a complexity of interesting features and exciting solvable mathematical models. In the present article, we have studied the one-dimensional negative Laplacian decorated with two equally weighted $\delta'$-interactions symmetrically distributed with respect to the origin. 

First of all, the need for a proper self-adjoint determination of the Hamiltonian, which implies the use of techniques such as renormalisation, is to be stressed. This is far from being trivial since it uses a determination of the resolvent of the self-adjoint operator as a norm resolvent limit of the resolvents of a net of approximating Hamiltonians. 

The use of the resolvent solves the eigenvalue problem for the studied Hamiltonian. In particular, we have shown that this model has two eigenvalues and have studied their behaviour  as functions of $x_0$. We have shown the existence of resonances and the absence of other scattering features such as anti-bound or redundant states. 

We have also taken the limit as $x_0 \to 0^+$ and shown that in this limit the two perturbations merge smoothly yielding a single point perturbation with double strength. We have compared this model with others investigated in the past.

\section*{Acknowledgments}

First of all, we wish to thank the anonymous referees whose constructive criticism has led to the overall improvement of our manuscript. This research was supported by Spanish MCIN with funding from European Union Next Generation EU (PRTRC17.I1) and Consejeria de Educacion from JCyL through QCAYLE project, as well as MCIN projects PID2020-113406GB-I00 and RED2022-134301-T. S. Fassari would like to thank Prof. F. Rinaldi as well as the other members of the Engineering Department of Marconi University (Rome) for their kind invitation to present the early stages of this work during the workshop ``Risultati recenti sulle Interazioni Puntuali in Meccanica Quantistica e loro applicazioni" (Recent results on point interactions in Quantum Mechanics and their applications) held at Marconi University on 30th November 2022. S. Fassari wishes to express his heartfelt thanks to Prof. Nieto and Prof. Gadella for making his stay at their institution (Department of Theoretical Physics, Atomic Physics and Optics, University of Valladolid, Spain) possible during the second half of April 2023 through the aforementioned funding sources, as well as to all the other members of the Department for their warm hospitality.

\section*{Data Availability Statement}

No Data associated in the manuscript.


\begin{thebibliography}{99}

\bibitem{AGHH} Albeverio S, Gesztesy F, H{\o}egh-Krohn R and Holden H 2004
{\it Solvable Models in Quantum Mechanics} (Providence, RI: AMS Chelsea
Series)

\bibitem{AK} Albeverio S and Kurasov P 2000 {\it Singular Perturbations of
Differential Operators} ({\it Lecture Note Series vol 271}
(Cambridge: London Mathematical Society)

\bibitem{BR}  Belloni M and  Robinett R W 2014 {\it Phys. Rep.} {\bf 540} 25--122

\bibitem{DO}  Demkov Yu N and V. N. Ostrovskii V N 1998 {\it Zero-Range
Potentials and Their Applications in Atomic Physics}
(New York: Plenum Press)

\bibitem{ZZ}  Zolotaryuk A V and  Zolotaryuk Y 2015 {\it J. Phys. A} {\bf 48} 035302

\bibitem{ZZ1} Zolotaryuk A V and  Zolotaryuk Y 2015 {\it Phys. Lett. A} {\bf 379} 511--517

\bibitem{FI}  Fassari S and  Inglese G 1994 {\it Helv. Phys. Acta} {\bf 67} 650--659

\bibitem{FI1}  Fassari S and  Inglese G 1997 {\it Helv. Phys. Acta} {\bf  70} 858--865

\bibitem{Z}  Zolotaryuk A V  2017 {\it J. Phys. A: Math.  Theor.} {\bf 50}  225303

\bibitem{FGGN} Fassari S, Gadella M, Glasser M L and Nieto L M 2018 {\it Ann. Phys. N.Y.} {\bf 389} 48--62

\bibitem{MMM}  Mu\~noz Castaneda J M,  Mateos Guilarte J and  Mosquera A M 2013 {\it Phys. Rev. D} {\bf 87}  105020

\bibitem{AGM}  Asorey M,  Garcia-Alvarez D and  Mu\~noz Casta\~neda J M 2006 {\it J. Phys. A} {\bf 39} 6127--6136

\bibitem{AM} Asorey M and Mu\~noz Casta\~neda J M 2013 {\it Nucl. Phys. B} {\bf 874} 852--876.

\bibitem{MKB} Mu\~noz Casta\~neda J M, Kirsten K and Bordag M 2015 {\it Lett. Math. Phys.} {\bf 105} 523--549

\bibitem{MGMC} Mateos Guilarte J and Mu\~noz Casta\~neda J M 2011 {\it Int. J. Theor. Phys.} {\bf 50} 2227--2241

\bibitem{BM} Bordag M and Mu\~noz Casta\~neda J M 2012 {\it J. Phys. A} {\bf 45} 374012

\bibitem{BOR} Bordag M 2014 {\it Phys. Rev. D} {\bf 89} 125015

\bibitem{UTM}  Uncu H,  Tarhan D,  M\"ustecapl{\i}o\u{g}lu \"O E 2007 {\it Phys. Rev. A} {\bf  76}  013618

\bibitem{APST}  Avakian M P,  Pogosyan G S,  Sissakian A N and  Ter-Antonyan V M 1987 {\it Phys. Lett. A}  {\bf 124} 233--236

\bibitem{DE}  Demiralp E 2005 {\it J. Phys. A: Math. Gen.} {\bf 22} 4783--4793

\bibitem{GMMN}  Gadella M,  Mateos-Guilarte J,  Mu\~noz-Casta\~neda J M,  Nieto L M and Santamar\'ia-Sanz L 2020 {\it Eur. Phys. J. Plus} {\bf 135} 786

\bibitem{ZZ2} Zolotaryuk A V and  Zolotaryuk Y 2020 {\it Low Temp. Phys.} {\bf 46}  779--785

\bibitem{Z1} Zolotaryuk A V, Christiansen, P L and Iermakova S V 2007 {\it J. Phys. A: Math. Theor.} {\bf 40}5443--5457

\bibitem{Z2} Christiansen P L,  Arnbak H C, Zolotaryuk A V, Ermakov V N and Gaididei Y B 2003 {\it J. Phys. A: Math.  Theor.}  {\bf 36} 7589-7600

\bibitem{EK}  Espinosa M G and   Kielanowski P 2008 {\it J. Phys.: Conf. Ser.} {\bf  128} 012037

\bibitem{EGU} Erman F,  Gadella M,  Uncu H 2017 {\it Phys. Rev. D} {\bf 95} 045004

\bibitem{EGU1} Erman F,  Gadella M, Tunal{\i} S and  Uncu H 2017 {\it Eur. Phys. J. Plus} {\bf 132} 352

\bibitem{EGU2} Erman F,  Gadella M,  Uncu H 2018 {\it EJP} {\bf 39} 035403

\bibitem{AGLM} Alvarez J J,  Gadella M,  Lara L P,  Maldonado-Villamizar F H 2013 {\it Phys. Lett. A} {\bf 377} 2510--2519

\bibitem{KU}  Kurasov P 1996 {\it J. Math. Ann. Appl.} {\bf  201} 297--323

\bibitem{KP}  Kulinskii V L  and   Panchenko D Y 2015 {\it Physica B Condens. Matter} {\bf 472} 78--83

\bibitem{KP1}  Kulinskii V L  and   Panchenko D Y 2019 {\it Ann. Phys.} {\bf 404} 47--56

\bibitem{AK1} Albeverio S and Kurasov P 1997 {\it Lett. Math. Phys.} {\bf 41} 79--92

\bibitem{HSW}  Al-Hashimi M H,  Shalaby A M, and Wiese U J 2014 {\it Phys. Rev. D} {\bf 89} 125023

\bibitem{HS} Al-Hashimi M H and  Shalaby A M 2015 {\it Phys. Rev. D} {\bf 92} 025043

\bibitem{FI2}  Fassari S and  Inglese G 1996 {\it Helv. Phys. Acta} {\bf 69}  130--140

\bibitem{BG} Br\"uning J, Geyler V and Lobanov I 2004 {\it J. Math. Phys.} {\bf 45}  1267--1290

\bibitem{AFR}  Albeverio S,  Fassari S and Rinaldi F 2016 {\it Nanosyst.: Phys. Chem. Math.} {\bf 7} 268--289

\bibitem{AFR1}  Albeverio S,  Fassari S and Rinaldi F 2016 {\it Nanosyst.: Phys. Chem. Math.} {\bf 7} 803--815

\bibitem{E} Erman F 2017 {\it Commun. Theor. Phys.} {\bf 68} 313--316

\bibitem{EST} Erman F, Seymen S and Turgut O T 2022  {\it J. Math. Phys.} {\bf 63} 123505

\bibitem{LL} Lieb E H and Loss M 2001 {\it Analysis} (Providence, RI: AMS)

\bibitem{K} Klaus M 1982 {\it Helv. Phys. Act.} {\bf 55} 413--419

\bibitem{K1} Klaus M 1977 {\it Ann. Phys.} {\bf 108} 288--300

\bibitem{RSI} Reed M, Simon, B 1972 {\it  Methods in Modern Mathematical Physics: Functional Analysis} (New York: Academic Press)

\bibitem{Scharf} Scharf G  1989 {\it Finite Quantum Electrodynamics} (Berlin-Heidelberg: Springer Verlag)

\bibitem{Tadeu} Cal\c{c}ada M, Lunardi JT, Manzoni LA and Monteiro W 2014 {\it Front. Phys.} {\bf2}:23

\bibitem{RSII} Reed M, Simon, B 1975 {\it  Methods in Modern Mathematical Physics: Fourier Analysis. Self Adjointness} (New York: Academic Press)

\bibitem{DONSIMON} Simon B 1971 {\it Quantum Mechanics for Hamiltonians Defined as Quadratic Forms} (Princeton: Princeton University Press)

\bibitem{GLP} Grosse H, Langmann E and Paufler C 2004 \textit{ J. Phys. A: Math Gen.} {\bf 16} 4579–92

\bibitem{AFR4} Albeverio S,  Fassari S and  Rinaldi F 2013 \textit{J. Phys. A: Math. Theor.} {\bf 46} 385305

\bibitem{Qc} Schmidtke H-H 1987 {\it Quantenchemie} (Weinheim: VCH) (in German)

\bibitem{BB} Byers Brown W, Steiner E 1966 {\it J. Chem. Phys.} {\bf  44} 3934

 \bibitem{KH2+} Klaus M 1983 {\it J. Phys. A Math. Gen.} {\bf 16} 2709--2720

\bibitem{KDW} Klaus M 1981 {\it Ann. Inst. Henri Poincar\'e, A Phys. th\'eor.} {\bf 34} 405--417

\bibitem{AFR3} Albeverio S,  Fassari S and  Rinaldi F 2017 {\it Nanosyst.: Phys. Chem. Math.}  {\bf 8} 153--157

\bibitem{FST}  Figari R, Saberbaghi H, Teta A. 2023 arXiv:2306.10292 [math-ph]

\bibitem{FPR}  Fassari S, Popov I and  Rinaldi F 2020 \textit{Phys. Scr.} {\bf 95} 075209

\bibitem{AFR2}  Albeverio S,  Fassari S and  Rinaldi F 2015 \textit{J. Phys. A: Math. Theor.}  {\bf 48} 185301.

\bibitem{ADK} Albeverio S, Dabrowski L and Kurasov P 1998 {\it Lett. Math. Phys.} {\bf 45} 33--47

\bibitem {CAFEPO} Cacciapuoti C, Fermi D and Posilicano A 2022 {\it Reviews in Mathematical Physics} {\bf 34} 6 2250015 

\bibitem{FR} Fassari S and  Rinaldi F 2009 \textit{Rep. Math. Phys.} {\bf 64} 367--393

\bibitem{KS} Kovařík H and Sacchetti A 2010  \textit{J. Phys. A: Math. Theor.} {\bf 43} 155205

\bibitem{S} Sacchetti A 2016  \textit{J. Phys. A: Math. Theor.} {\bf 49} 175301

\bibitem{SCorr} Sacchetti A 2016  \textit{J. Phys. A: Math. Theor.} {\bf 49} 439501

\bibitem{KLA} Klaus M 1979 \textit{Helv. Phys. Acta} {\bf 52} 223--229

\bibitem{FA} Fassari S 1995 \textit{Helv. Phys. Acta} {\bf 68} 121--125

\bibitem{FGNR} Fassari S, Gadella M, Nieto LM and Rinaldi F 2017 {\it Acta Polytech.} {\bf 57} 385--390

\bibitem{Cav} Cavalcanti R.M. 1999 \textit{Rev. Bras. Ens. Fis.} {\bf 21} 336

\bibitem{EX} Exner P, Neidhardt H, Zagrebnov VA 2001  \textit{Commun. Math. Phys.} {\bf 224} 593612

\bibitem{AFR5} Albeverio S,  Fassari S and  Rinaldi F 2016 \textit{J. Phys. A: Math. Theor.} {\bf 49} 025302

\bibitem{FGNR1} Fassari S, Gadella M, Nieto LM and Rinaldi F 2021 \textit{Eur. Phys. J. Plus} {\bf 136} 673

\bibitem{FR1} Fassari S and  Rinaldi F 2012 \textit{Rep. Math. Phys.} {\bf 69} 353--370

\bibitem{NUS} Nussenzveig HM, {\it Causality and Dispersion Relations} 1972 (New York: Academic Press)

\bibitem{KKH} Kukulin V I, Krasnopolski V M, Hor\'a\v{c}ek J 1989 {\it Theory of Resonances. Principles and Applications} (Dordrecht: Kluwer)

\bibitem{ID} Dassie AC, Gerdau F, Gonz\'alez FJ, Moyano M, Id Betan RM 2022 {it Am. J. Phys.} {\bf 90} 118--125

\bibitem{GKN} Gadella M, Kuru {\c S}, Negro J 2017 {\it Ann. Phys.} {\bf 379} 86--101

\bibitem{MA} Ma ST 1946 {\it Phys. Rev.} {\bf 69} 668--668

\bibitem{MM} Moroz A, Miroshnichenko AE 2019 {\it EPL} {\bf 126} 30003

\bibitem{GHKN} Gadella M, Hern\'andez-Ortega A, Kuru {\c S}, Negro J 2020 \textit{Eur. Phys. J. Plus} {\bf 135} 822

\bibitem{TraceId} Simon B 1979 {\it Trace Ideals and Their Applications} (Cambridge: Cambridge University Press)

\bibitem{K3} Klaus M 1980 {\it J. Phys. A: Math. Theor.} {\bf13} L205--L208

\bibitem{FGGNR} Fassari S, Gadella M, Glasser, ML, Nieto LM, Rinaldi F 2018 {\it Nanosyst.: Phys. Chem. Math.} {\bf 9} 179--186

\bibitem{RSIV} Reed M, Simon B 1978 {\it Methods in Modern Mathematical Physics: Analysis of Operators} (New York: Academic Press)












\end{thebibliography}
\end{document}